\documentclass[twocolumn]{aastex62}
\usepackage[utf8]{inputenc}
\usepackage{amsmath}
\usepackage{graphicx}
\usepackage{subfigure}
\usepackage{lipsum}
\usepackage{lineno}
\submitjournal{AJ}
\accepted{May 16, 2022}

\newcommand{\teff}{\mbox{$\rm T_{\rm eff}$}}

\newcommand{\logg}{\mbox{$\log g$}}

\newcommand{\vsini}{\mbox{$v \sin i$}}

\newcommand{\kms}{\mbox{km\,s$^{-1}$}}

\newcommand{\rsun}{R\ensuremath{_\odot}}
\newcommand{\msun}{M\ensuremath{_\odot}}

\newcommand{\rstar}{\ensuremath{R_\star}}
\newcommand{\mstar}{\ensuremath{M_\star}}

\newcommand{\asini}{\ensuremath{a \sin{i}}}

\begin{document}

\title{Modeling Stellar Surface Features on Subgiant Star with an M-dwarf Companion}
\author[0000-0003-2435-130X]{Maria C. Schutte}
\email{maria.schutte-1@ou.edu}
\affiliation{Homer L. Dodge Department of Physics and Astronomy, University of Oklahoma, 440 W. Brooks Street, Norman, OK 73019, USA}

\author{Leslie Hebb}
\affiliation{Department of Physics, Hobart and William Smith Colleges, 20 St. Clair Street, Geneva, NY 14456, USA}

\author{Simon Lowry}
\affiliation{Homer L. Dodge Department of Physics and Astronomy, University of Oklahoma, 440 W. Brooks Street, Norman, OK 73019, USA}

\author{John Wisniewski}
\affiliation{Homer L. Dodge Department of Physics and Astronomy, University of Oklahoma, 440 W. Brooks Street, Norman, OK 73019, USA}

\author[0000-0002-6629-4182]{Suzanne L. Hawley}
\affiliation{Department of Astronomy, University of Washington, 3910 15th Ave NE, Seattle, WA 98195}

\author[0000-0001-9596-7983]{Suvrath Mahadevan}
\affiliation{Department of Astronomy and Astrophysics, Penn State University, 525 Davey Laboratory, University Park, PA 16802}
\affiliation{Center for Exoplanets \& Habitable Worlds, Penn State University, 525 Davey Laboratory, University Park, PA 16802}

\author[0000-0003-2528-3409]{Brett M. Morris}
\affiliation{Center for Space and Habitability, Gesellsschaftstrasse 6, 3012 Bern, Switzerland}

\author{Paul Robertson}
\affiliation{Department of Physics and Astronomy, University of California - Irvine, 4129 Frederick Reines Hall, Irvine, CA 92697}

\author{Graeme Rohn}
\affiliation{Department of Physics, State University of New York at Cortland,
22 Graham Ave, Cortland, NY 13045}

\author{Gudmundur Stefansson}
\affiliation{Department of Astrophysical Sciences, Princeton University, 4 Ivy Lane, Princeton, NJ 08544}
\affiliation{Henry Norris Russell Fellow}

\begin{abstract}
Understanding magnetic activity on the surface of stars other than the Sun is important for exoplanet analyses to properly characterize an exoplanet's atmosphere and to further characterize stellar activity on a wide range of stars. Modeling stellar surface features of a variety of spectral types and rotation rates are key to understanding of the magnetic activity of these stars. Using data from \textit{Kepler}, we use the starspot modeling program STarSPot (\texttt{STSP}) to measure the position and size of spots for KOI-340 which is an eclipsing binary consisting of a subgiant star ($\teff = 5593 \pm 27 K; \rstar = 1.98 \pm 0.05 \rsun$) with an M-dwarf companion ($\mstar = 0.214 \pm 0.006 \msun$). \texttt{STSP} uses a novel technique to measure the spot positions and radii by using the transiting secondary to study and model individual active regions on the stellar surface using high-precision photometry. We find the average size of spot features on KOI-340's primary is $\sim$10\% the radius of the star, i.e. two times larger than the mean size of Solar-maximum sunspots. The spots on KOI-340 are present at every longitude and show possible signs of differential rotation. The minimum fractional spotted area of KOI-340's primary is $2\substack{+12\\ -2} \%$ while the spotted area of the Sun is at most 0.2\%. One transit of KOI-340 shows a signal in the transit consistent with a plage; this plage occurs right before a dark spot indicating the plage and spot might be co-located on the surface of the star.
\end{abstract}

\section{Introduction}\label{sec:intro}

 Sunspots, the locations of where magnetic fields tangle and erupt to the surface of the Sun, have been observed and studied since the late 19th century. 
 They are known to form in groups and have a non-uniform temperature with a darker cooler umbra surrounded by a slightly warmer, brighter penumbra. Sunspot groups are complex in nature, and their size and overall complexity play a significant role in the underlying magnetic activity level \citep{zirin1998}. The overall sizes of sunspot groups vary on the eleven year solar cycle with groups range in radius from $\sim 0.3~\%$ of the Sun during a typical Solar minimum period and on average $\sim 7~\%$  during a typical solar maximum period with the largest sunspot on record having a radius of 11\% the radius of the Sun \citep{newton1955}. Sunspots also grow and decay over time, though for individual spots the decay rate appears to be constant no matter the area of the spot. The number of sunspots in each group increases as the area of the group gets larger, meaning there are more likely to be many small spots within a group rather than one large spot \citep{hathaway2008}.

Detailed studies of starspots, the stellar equivalent of sunspots, are key to understanding the magnetic activity on the surface of stars. These studies expose key underlying magnetic features which can provide valuable constraints on stellar dynamos \citep{bergyugina2005}. Stellar activity that is similar to the Sun can be seen on M-dwarfs, including a subset of active M-dwarfs that exhibit significantly more activity and more energetic flares than seen on the Sun (see e.g. \citealt{bergyugina2005}). Stellar activity on other spectral types can be found as well through studying the brightness fluctuations of the stars over time \citep{kron1952}. For solar type stars, there is also a distinct connection between the rotation period of a star and the magnetic activity level, with faster rotating late-type stars exhibiting stronger activity \citep{skumanich1972}. Thus, a late-type star with a fast rotational period will be much more active than a slowly rotating star of a similar temperature. Furthermore, giant stars in short-period binary systems, e.g. RS CVn systems, can also be much more active than their younger, main sequence counterparts \citep{oneal2004}.

Analyses of the spotted areas of stars is also key to studying planet atmospheres using transmission spectroscopy as stellar activity is a major source of contamination for these observations \citep{Rackham2022}. When considering the effect of spots on the surfaces of stars in transmission spectroscopy, the overall fraction of the star that is covered in spots is a key parameter, and \citet{pont2007} showed that for even a total spotted area of 1\%, starspots would be the dominant source of uncertainty for a transmission spectrum of an exoplanet. \citet{rackham2018} also showed that spot and faculae covering fractions for M-dwarfs generally underestimate the stellar contamination, and with realistic stellar contamination levels, the resulting transit depth effects can be up to 10 times that of planetary atmospheric features. This has been shown with the M dwarf K2-18 where \citet{barclay2021} found that stellar surface brightness variations could explain the inferred detection of a water absorption feature on its sub-Neptune habitable zone planet's atmosphere found by \citet{benneke2019}.  

With some active stars having spotted areas of up to 40\% \citep{oneal2001}, the characterization of the activity on all stars is imperative to correctly understanding the transmission spectra of exoplanets around all stars. At the high levels of spectrophotometric precision that missions like JWST and ARIEL will provide, this is now a critical effect to understand and mitigate \citep{Rackham2022}. The Pandora SmallSat mission will also allow for simultaneous visible time-series photometry and near-IR spectroscopy of exoplanet targets to understand and mitigate the effect of stellar activity on exoplanet atmospheres \citep{quintana2021}.

There are many different techniques that have been developed to study starspots including Doppler imaging \citep{vogt1987}, observations of molecular lines \citep{oneal1996}, spectropolarimetry \citep{donati1997}, and long-term photometric observations \citep{morris2017}. Each individual technique is important because each one tells us about various aspects of starspots such as their temperature from molecular line observations, differential rotation with Doppler imaging, and stellar activity evolution from photometric observations \citep{bergyugina2005}. Additionally, high precision transit photometry can allow one to spatially resolve starspots, using the transit (whether planetary or stellar) as a knife-edge probe of the star. While the transit is occurring, the overall flux of the host star is reduced. If the companion crosses in front of a spot (or a plage), there will be a signature positive (or negative) bump during the in-transit part of the light curve. Both spectroscopy and transit photometry provide insight into the overall spotted area of a star, i.e. the filling or covering factor.  Spectroscopic observations probe the net amount of spatially unresolved spots, whereas transit photometry can spatially resolve spots along the transit chord blocked by the companion during transit.

\citet{wolter2009} were the first to use this novel technique to map a starspot on the surface of CoRoT-2 were the spot was occulted during a planetary transit. \citet{morris2017} and \citet{netto2020} then applied this novel photometric technique to \textit{Kepler} satellite data of the K-dwarf star, HAT-P-11, and the young solar analogue, Kepler-63, respectively.  
HAT-P-11 has a similar rotation period to the Sun at $\sim~$29 days, but as it is a cooler star, it has a deeper convection zone which could lead to higher levels of activity than seen in the Sun \citep{morris2017}. The deeper convection zone of cooler stars leads to more turbulence in the star and thus more activity \citep{bergyugina2005}. \citet{netto2020} found that 
Kepler-63 has two bands of spots in its Northern hemisphere with larger spots closer to the equator and pole. Kepler-63's fast rotation rate ($\sim~$5.4 day period) could explain why it has larger spots near the equator and pole of the star \citep{netto2020}.  Both \citet{morris2017} and \citet{netto2020} were able to study distribution of spot latitudes and the change in the spot latitudes over time because both planets have nearly polar orbits. \citet{netto2020} found little evidence of differential rotational in Kepler-63, indicating that it rotates almost as a solid body. 



\begin{figure}[h]
    \centering
    \begin{minipage}{0.48\textwidth}
        \centering
        \includegraphics[width=\textwidth]{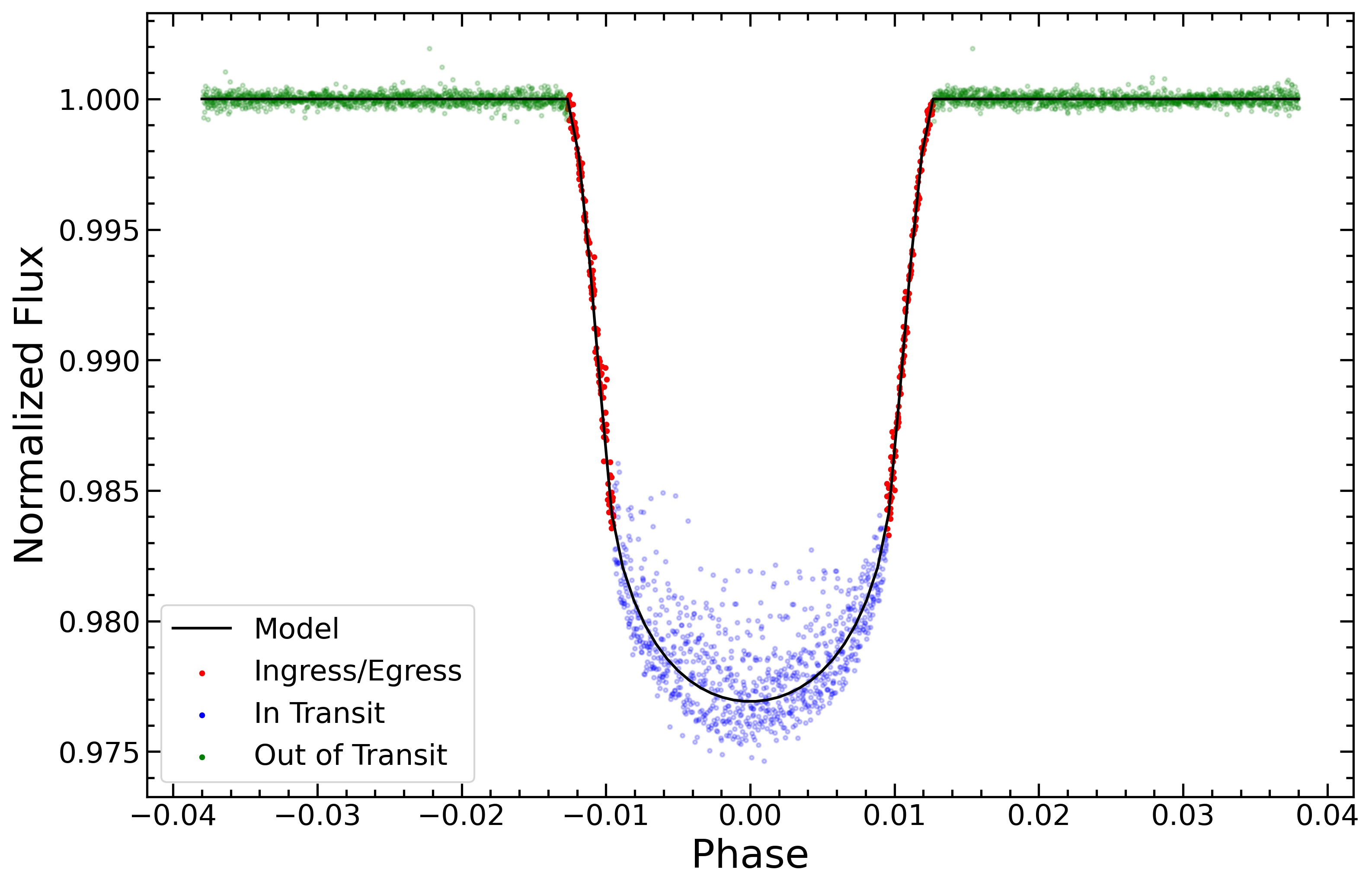} 
    \end{minipage}%
    \begin{minipage}{0.48\textwidth}
        \centering
        \includegraphics[width=\textwidth]{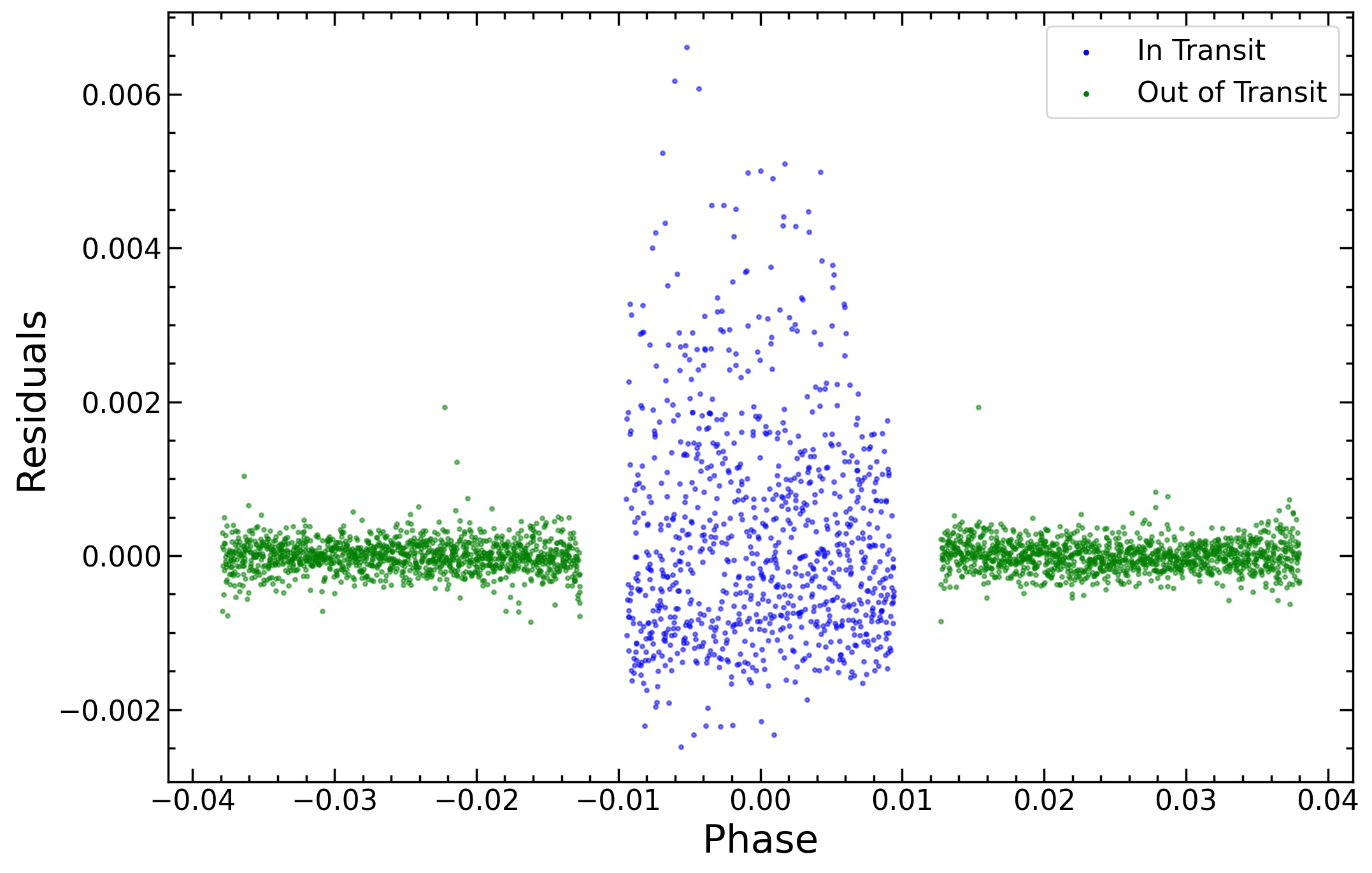} 
        \caption{Top: Normalized transit light curves of KOI-340 from the \textit{Kepler} DR25 \citep{thompson2018} data release are shown as a function of phase.  The out-of-transit data is displayed in green, the ingress and egress is in red, and the measurements made when the companion is in-transit (between the second and third contact points) are in blue. The black line model is generated by using the DR25 parameters for KOI-340 with the \texttt{batman} software \citep{kreidberg2015}.  Bottom: Residual \textit{Kepler} transit light curves of KOI-340 compared to the model generated from the parameters from the DR25 \citep{thompson2018} data release versus phase.  The out-of-transit data is in green, and the measurements made between the second and third contact points are in blue. The significant increase in in-transit scatter compared to the immediately adjacent out-of-transit data are the signature of the secondary star occulting starspots on the surface of the primary star. We have chosen not to plot the ingress and egress points to highlight the difference between the in-transit and out-of-transit points.}
    \end{minipage}
    \label{fig:kep_lcs}
\end{figure}

We identify KOI-340 (KIC 10616571, TIC 273376221, 2MASS J19503952+4748050) as an eclipsing binary system consisting of a G subgaint primary and an M dwarf companion that exhibits in-transit starspot crossing features in the original long cadence \textit{Kepler} data. 
Figure~\ref{fig:kep_lcs} shows all 38 normalized transits of KOI-340 overplotted by the model generated from the DR25 data release \citep{thompson2018}. The residuals compared to that model plotted in Figure~\ref{fig:kep_lcs}.  These residuals exhibit a significant increase in in-transit scatter compared to the immediately adjacent out-of-transit data, which is the signature of variations in surface brightness (i.e.\ active regions) on the primary star being occulted by the secondary star during the transit.  

In this paper, we model the \textit{Kepler} long cadence light curves of KOI-340 to characterize the starspots on the primary star.  In Section \ref{sec:params}, we present updated physical and orbital parameters of the eclipsing system to determine the unspotted transit light curve. In Section \ref{sec:STSP}, we describe the modeling of all \textit{Kepler} transits showing in-transit starspot crossing features to derive the properties of the spots.   In Section \ref{sec:res_dis}, we discuss the implications of the derived spot properties and put KOI-340 into the broader context of known objects with starspot measurements. Finally, in Section \ref{sec:conc}, we summarize the main conclusions of the paper. 

\section{Physical and Orbital Properties of KOI-340}\label{sec:params}



 KOI-340  is a highly eccentric eclipsing binary consisting of a G subgiant primary star with an M-dwarf companion in a 23.67~day orbit.
\citet{santerne2012} were the first to conclude that the system was a single-line spectroscopic binary based on two radial velocity (RV) measurements made at quadrature with the SOPHIE instrument.  An additional 21 RV measurements obtained at eight different phases with the high resolution spectrograph, CAFE, on the 2.2-meter telescope at Calar Alto Observatory \citep{aceituno2013} were modeled along with the primary and secondary transits from \textit{Kepler} \citep{lillobox2015} to derive the orbital parameters and determine this is a highly eccentric ($e  = 0.513 \pm 0.005$) eclipsing system
with a mass ratio, $q = 0.20 \pm 0.05$.

 \citet{brewer2018} present the spectroscopic stellar parameters, $\teff$, $\logg$, \vsini\, and metallicity, for KOI-340 as determined by fitting its stellar spectrum with synthetic spectra, and fit isochrones to obtain a mass and radius estimate for the primary star. Using these techniques, the authors find the following stellar parameters for the primary star: $\teff$ = 5593$\pm 27$~K, $\logg$ = 3.96$\pm 0.05$, [M/H] = 0.28$\pm 0.01$, $\vsini$ = 6.9$\pm 0.5$ \kms, $\mstar$~=~1.21$^{+0.04}_{-0.03}~\msun$, and $\rstar$~=~1.89$~\rsun$.   
Furthermore, \citet{mcquillan2013} analyzed 10 months of \textit{Kepler} data to measure the rotation period of 1570 objects. KOI-340 was included in this analysis, and \citet{mcquillan2013} found a rotational period for KOI-340 of $12.942 \pm 0.018$ days. 


The transit model presented in the DR25 \textit{Kepler} data release and shown as the black line in Figure~\ref{fig:kep_lcs}  was derived from all the primary transits of KOI-340, most of which have starspot crossing features.  This results in a default model with a depth that is shallower than what would have been derived from unspotted transits. However, accurate characterization of the starspots depends on an underlying transit model shape that reflects the non-spotted stellar surface flux with limb darkening.  In addition, correct characterization of the longitude and latitude of the starspots on the surface depends on the most accurate orbital properties of the two-body system and knowledge of the stellar rotation period and tilt of the spin axis of the star.    
Therefore, we derive updated values for the physical and orbital properties of the KOI-340 system using only a subset of transits that show little or no in-transit spots, while also incorporating in the fit the 21 existing radial velocity measurements from \cite{lillobox2015} and all the secondary transits in the \textit{Kepler} light curve to fully constrain the eccentric orbit.  

The primary and secondary transits are normalized while preserving the transit depth in the presence of out-of-transit variability using the technique discussed in \cite{morris2017}.  We normalize all primary and secondary transits using the following steps: 
\begin{itemize}
    \item Fitting and subtracting a second-order polynomial from the out-of-transit fluxes within 3 hours of each transit.
    \item Add the peak quarterly flux to each detrended transit (which approximates the unspotted brightness of the star)
    \item Divide the fluxes by that same peak value
\end{itemize}
This technique removes trends in flux due to stellar variability and normalizes the out-of-transit fluxes to near-unity, while maintaining a uniform transit depth over all transits.

\begin{figure}[h]
    \centering
    \begin{minipage}{0.48\textwidth}
        \centering
        \includegraphics[width=\textwidth]{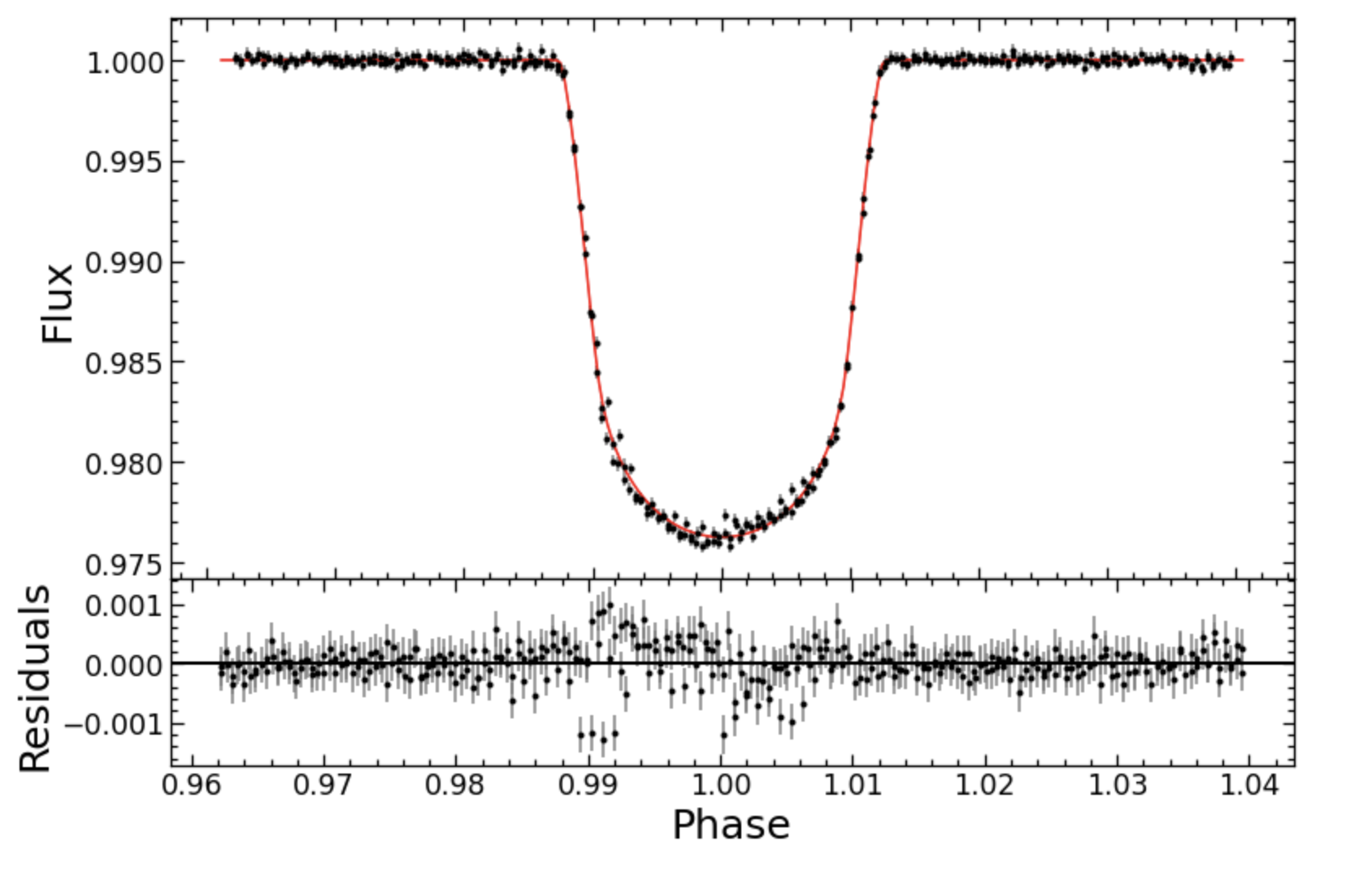} 
    \end{minipage}%
    \begin{minipage}{0.48\textwidth}
        \centering
        \includegraphics[width=\textwidth]{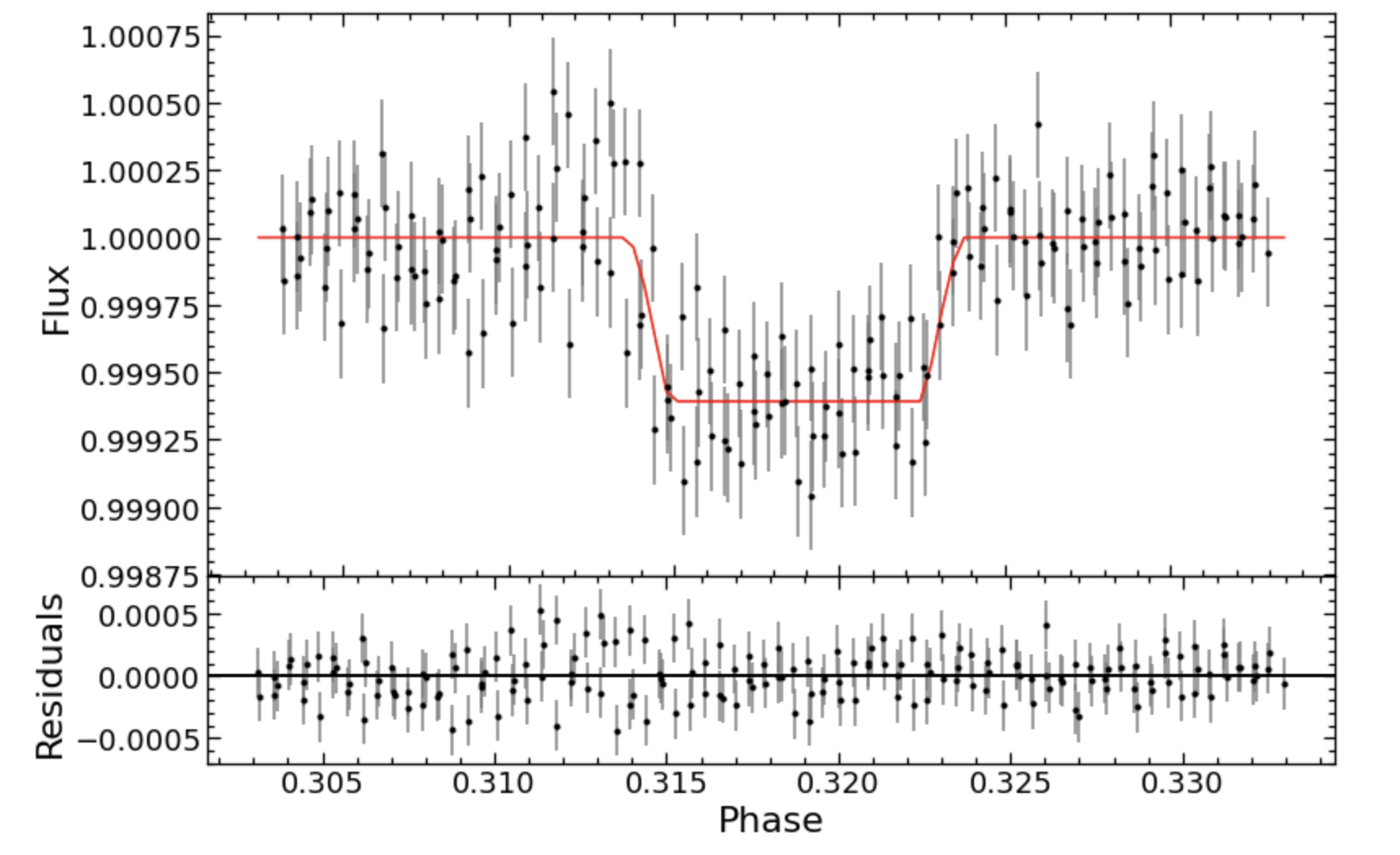} 
        \caption{Top: Four normalized transit light curves of KOI-340 which show little or no starspot crossing features.  The data are phase-folded with the period in Table~\ref{tab:tableparams} and plotted versus phase.  The red line \citet{mandel2002} model generated from the MCMC analysis of these primary transits along with the radial velocity measurements and secondary transits from \textit{Kepler}.  The depth of the transit derived from only unspotted transits is $\delta = 0.020164 \pm 0.000096$. Bottom: All normalized secondary transits observed in the long cadence \textit{Kepler} data of KOI-340 phase-folded with the ephemeris presented in Table~\ref{tab:tableparams} and plotted versus phase. The red line model is generated from the MCMC analysis described in Section~\ref{sec:params}.  The phase of the secondary transit occurs at 0.319 due to the eccentricity of the system.}
    \end{minipage}
    \label{fig:mcmc_params}
\end{figure}

\begin{figure}[h]
\centering
\includegraphics[width=0.48\textwidth]{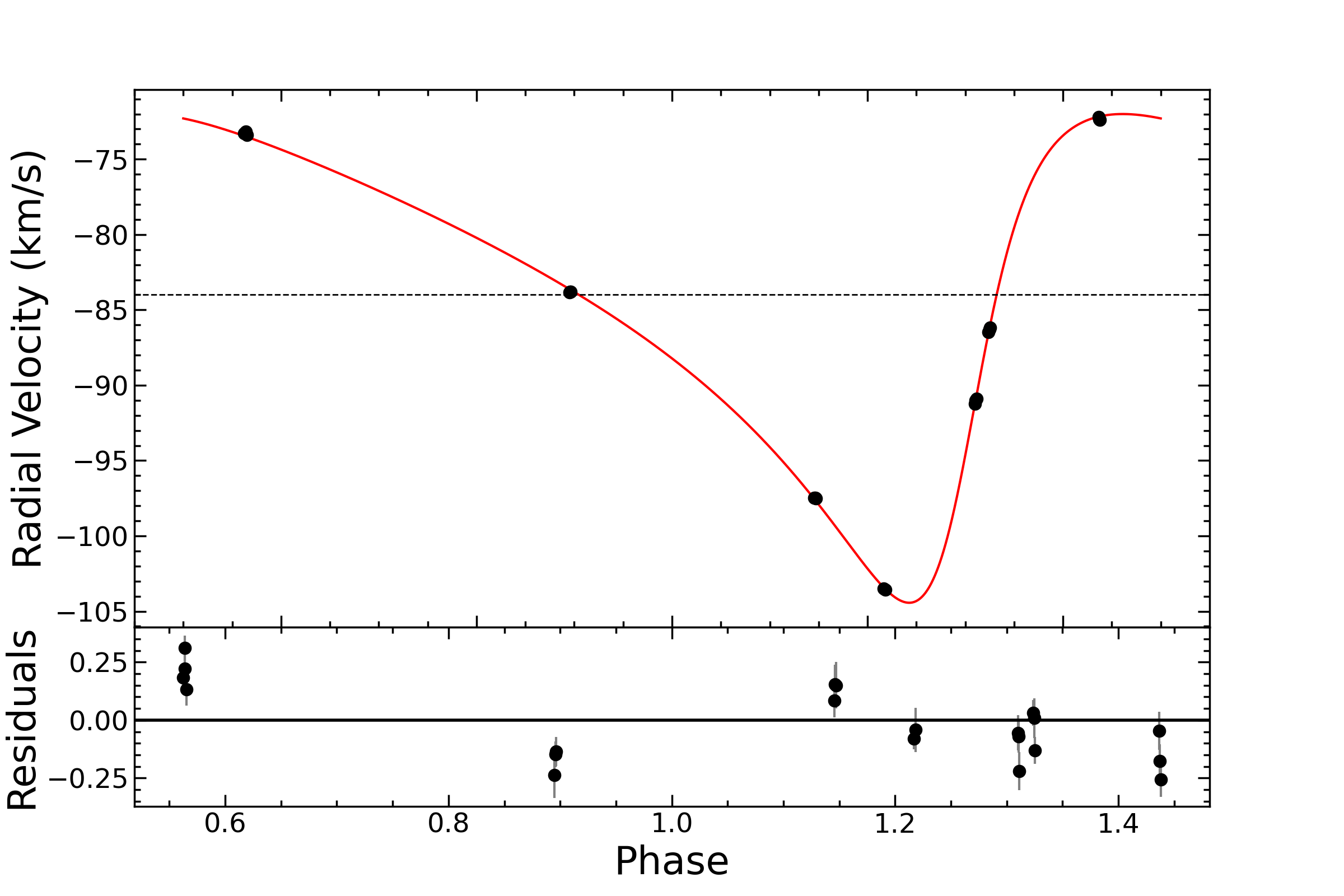}
\caption{Radial velocity points from \citet{lillobox2015} of KOI-340 phase-folded with the ephemeris presented in Table~\ref{tab:tableparams} and plotted versus phase.   The red line model is generated from the MCMC analysis described in Section~\ref{sec:params}.  }
\label{fig:rvplot}
\end{figure}

We then apply a Markov-Chain Monte Carlo (MCMC) analysis simultaneously to the normalized light curves and radial velocity points following the method described in \citet{cameron2007} and \citet{pollacco2008}.  Our light curve model adopts the analytic formulae presented in \citet{mandel2002} to describe the shape of the primary transit with limb darkening. As a single-lined eclipsing binary with a mass ratio, $q\sim 0.2$, we have chosen to use this well-tested code to characterize the precise shape of the unspotted primary transit and directly measure the orbital parameters that describe the position of the secondary companion. These parameters include the orbital period $P_{\rm orbital}$, the time of mid-primary transit $T_0$, the eccentricity $e$,
the argument of periastron $\omega$, the radial velocity semi-amplitude K$_1$,
the centre-of-mass velocity of the system $\gamma$, and depth of the secondary
transit ($\Delta$ F$_{\rm sec}$).   These parameters with their robust uncertainties are given in Table~\ref{tab:tableparams}.   While definitive masses and radii are beyond the scope of this paper, adopting a mass for the primary star from \citet{brewer2018} as mentioned above allows for analytically calculating the amplitude of the secondary radial velocity curve, K$_2$, the mass ratio, $q$, and the orbital separation, \asini\ by inverting the equations given in \citet{torres2010}.  These values are also provided in Table~\ref{tab:tableparams}.  Finally, the shape of the primary transit is accurately described by the \citet{mandel2002} model with a depth, $\delta =  0.020164 \pm 0.000096$, impact parameter, $b = 0.331 \pm 0.020$, and mean stellar density, $\rho_* = 0.208 \pm 0.018$ with theoretical four parameter limb darkening coefficients of $c_1 = 0.624, c_2 = -0.286, c_3 = 0.867, c_4 = -0.447$ which were determined using quasi-spherical PHOENIX model atmospheres \citep{claret}. The primary transit model and the secondary transit are shown in Figure \ref{fig:mcmc_params}. The radial velocity curve is shown in Figure \ref{fig:rvplot}.  

\begin{table}[h]
        \caption{Ephemeris and orbital properties of KOI-340 system}
        \begin{center}
        \begin{tabular}{lcc}
                \hline\\
                   & Value   & Units \\
                \hline\\
                T$_0$ & 6200.95698 $\pm$ 0.00025$^a$ & days \\
                P$_{\rm orbit}$ & 23.673113 $\pm$ 0.000011 &   days \\
                $e$ &   0.493 $\pm$ 0.019      &       \\
                $\omega$&      -122.0 $\pm$ 1.2   &       degrees \\
                K$_1$&  15.80 $\pm$ 0.40  & \kms \\
                $\gamma$       & -83.97 $\pm$ 0.02    & \kms \\
                $\Delta$ F$_{\rm sec}$&  0.00088 $\pm$ 0.00016 & \\
                K$_2$&  80.6 $\pm$ 1.0$^b$ &\kms \\
                q&$      0.20 \pm 0.05^b$ & \\
                $a \sin{i}$&$ 39.2 \pm 0.5^b $ & \rsun \\
                \hline\\
\end{tabular}
\end{center}
\label{tab:tableparams}
     {\footnotesize $^a$ Barycentric Julian Date -- 2\,450\,000 }\\
      {\footnotesize $^b$} Calculated assuming M~$=1.21$\msun \citep{brewer2018}
\end{table}

\begin{figure}[h]
\centering
\includegraphics[width=0.48\textwidth]{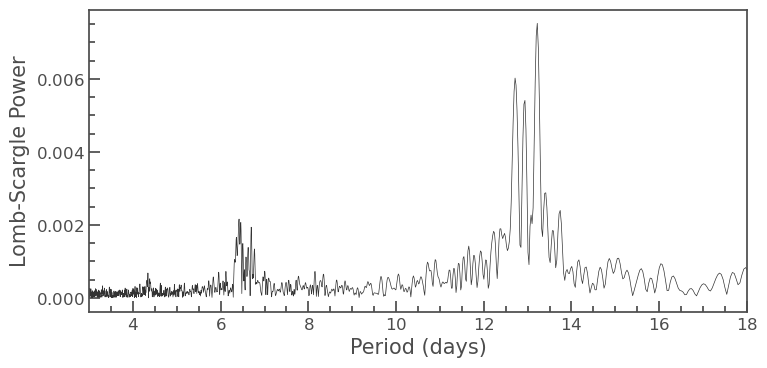}
\caption{The Lomb-Scargle Periodogram over all \textit{Kepler} data with primary transits removed. The period at maximum power suggests a rotational period for KOI-340 of 13.02 $\pm$ 0.97 days.}
\label{fig:LS}
\end{figure}

In addition to the orbital properties and shape of the transit, we 
measured the stellar rotation period and the tilt of the rotation axis 
out of the plane of the sky.
The period is measured from all available quarters of long cadence \textit{Kepler} light curves after eliminating the transits of the companion star.  
We run a Lomb-Scargle periodogram \citep{lombscargle} over the entire \textit{Kepler} data set with the primary transits removed from the data. The periodogram over all of the quarters is plotted in Figure \ref{fig:LS}. 

The Lomb-Scargle (LS) periodogram displays a maximum power at a period of 13.02 $\pm$ 0.97 days, which is consistent with the more robustly measured rotation period using the auto-correlation technique ($12.942~\pm~0.018$ days) as found by \citet{mcquillan2013}. If you instead run a periodogram on each individual \textit{Kepler} quarter, a range of rotational periods is obtained with the average period of those individual periodograms being 12.93 days with a standard deviation of 0.28 days which is also in agreement with the period found by \citet{mcquillan2013}. The relatively large range of measured rotation periods is likely due to real physical phenomena such as the stochastic emergence of spots at different phases and/or the differential rotation of the star leading to spots that emerge at different latitudes. \citet{suto2022} recently showed that multiple different peaks when examining the LS periodograms for each \textit{Kepler} quarter of an object may be used to put possible constraints on the differential rotation of the object. However, they do assume that the spots on the surface of the object do not evolve within one quarter ($\sim$ 90 days), which may not be the case for KOI-340 (see Section \ref{sec:long} for more details). As such, we leave further exploration of the quarter-to-quarter LS periodogram differences to future work. 

 A comparison between the measured \vsini\ and the \vsini\ calculated from the rotation period and stellar radius can be used to constrain the tilt of the rotation axis of the star.  The \vsini\ was measured by \cite{brewer2018} to be $6.9\pm 0.5$~\kms.  Adopting the radius, $\rstar~=~1.89 \pm 0.05$~\rsun\ determined from isochrone fitting in \citet{brewer2018} and using the rotation period derived by \citet{mcquillan2013}, we calculate the \vsini\ to be $7.40 \pm 0.03~\kms$. This value is consistent with a tilt of $0^\circ$ toward the observer and implies that the companion star will pass over approximately the same latitudes during every transit. Assuming the companion's orbit is aligned with the star's stellar rotation axis, i.e. $\lambda = 0^\circ$, \texttt{STSP} calculates this latitude as $-19.3^\circ$ based on the measured impact parameter of $b=0.331$. In the absence of an obliquity measurement, we assume the M dwarf companion's orbit is aligned with the host star so we can determine the latitude and longitude of the active regions location on the surface of the primary star. As shown in Section \ref{sec:long}, we do track spots over multiple orbits meaning the system is likely aligned. 



\section{Modeling Active Regions on KOI-340}\label{sec:STSP}

 We model the long cadence \textit{Kepler} light curves of 38 primary transits of KOI-340 using the modeling program, STarSPot (\texttt{STSP}) \citep{morris2017} to characterize the surface starspot features around the latitude of $-19.3^\circ$.  \texttt{STSP} is a C based program that models the surface brightness variations (i.e.\ starspots) on the primary star's photosphere in a two-body gravitationally bound eclipsing or transiting system.    

The parameters for the transit model derived above define the unspotted transit, but almost all of KOI-340's \textit{Kepler} transits contain evidence of stellar surface activity features, namely starspots and plages. Using this model as a starting point, we employ \texttt{STSP} to derive simulated spotted light curves of the primary transit of KOI-340 by adding a fixed number of spots each with a fixed contrast to the surface of the star. The contrast of the active regions is decided by the ratio of the integrated flux for the active region over the integrated flux for the star's effective photosphere relative to a certain bandpass.  We use a contrast equal to the average area-weighted contrast for sunspots ($c = 0.3$) for every spot \citep[and references therein]{morris2017}. We are modeling \textit{Kepler} light curves, so the contrast value we use is for the \textit{Kepler} bandpass. Given the flux of the secondary is around 2 orders of magnitude less than the primary, the effects modeled here are coming from the primary star.

Using an affine-invariant Markov Chain Monte Carlo (MCMC) \citep{foreman2013}, \texttt{STSP} optimizes the transit model by sampling different radii (R$_{\rm spot}/$R$_*$) and positions (latitude and longitude, $\theta$ and $\phi$ respectively) of each spot, ultimately adopting the model that produces the lowest $\chi^2$. 
We perform an initial MCMC run for every transit by starting 300 chains with one or two spots placed in random initial conditions on the star and allowing the spot configuration to evolve for an initial fixed time of approximately 4000 steps with each chain evolving independently. Then, we choose the chain with the lowest $\chi^2$ to be the best fit \texttt{STSP} model for that run.    
If the minimum $\chi^2$ solution to this run matches the spot features in the transit with a reduced $\chi^2$ less than 10, we consider this to be the final spot configuration for that transit. If the initial run appears to have the correct number of spots to match all the occultation features but not a sufficiently low reduced $\chi^2$, we continue running the MCMC optimizer for more time, re-starting from the last accepted step for all chains, until it reaches the required minimum $\chi^2$ (i.e. less than 10). We chose to consider our runs complete with a reduced $\chi^2$ of less than 10 as after running the transits for over 40,000 steps the $\chi^2$ values were not decreasing anymore and corresponded to a reduced $\chi^2$ of less than 10.

\begin{figure}[h]
    \centering
    \begin{minipage}{0.48\textwidth}
        \centering
        \includegraphics[width=\textwidth]{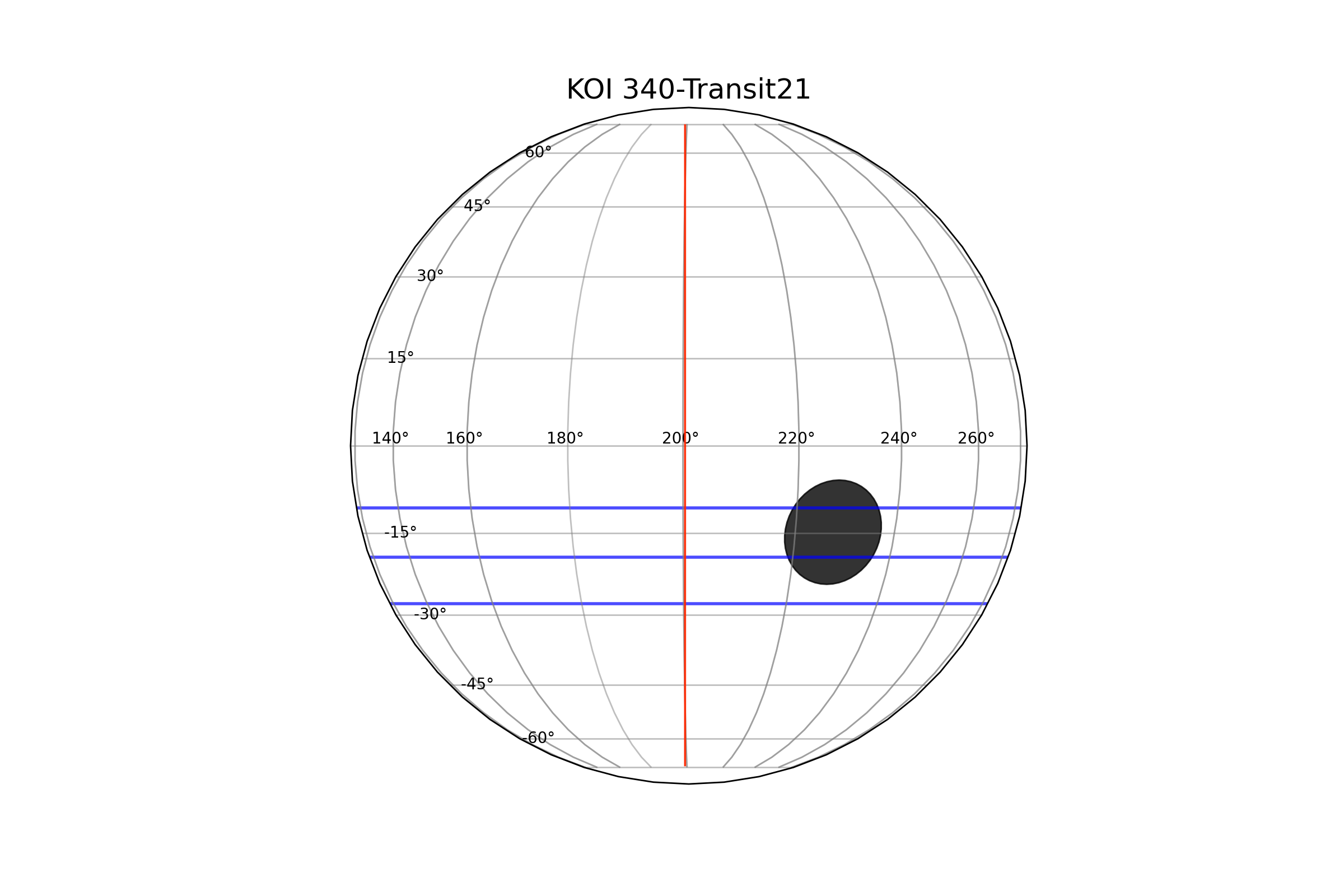} 
    \end{minipage}%
    \begin{minipage}{0.48\textwidth}
        \centering
        \includegraphics[width=\textwidth]{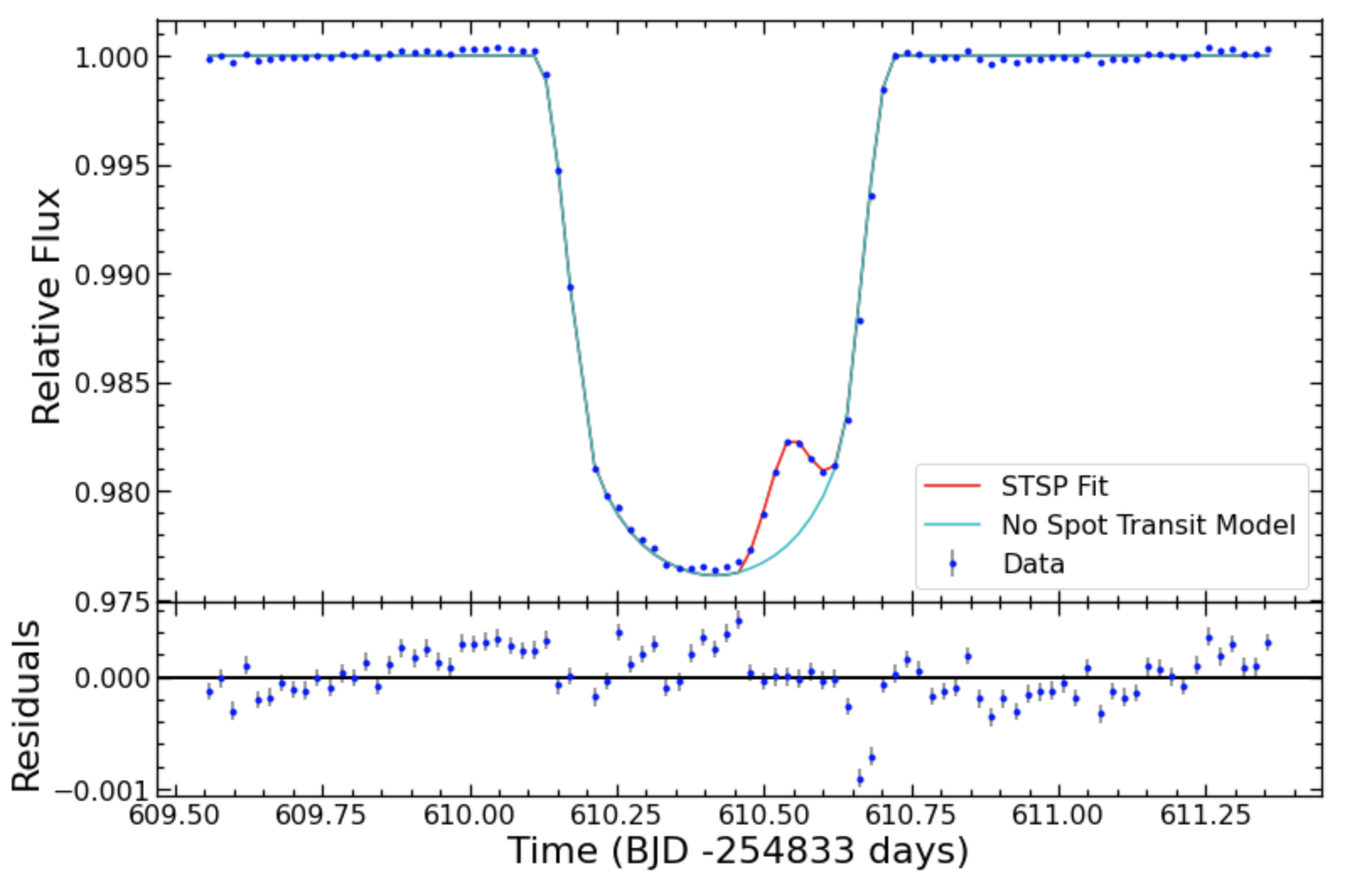} 
        \caption{Top: Plot of the surface of KOI-340 with the final spot groups shown as black filled circles along with the red line denoting the longitude of the star at mid-transit and with blue lines denoting the full extent of the transit path for the secondary object. Bottom: Light curve for final \texttt{STSP} fit (red line) along with the no spot model for KOI-340 (cyan line) for Transit 21. The residuals (model - data) are shown below the light curve with blue point with the error bars shown as light gray lines.}
    \end{minipage}
    \label{fig:T21}
\end{figure}

As an example, consisder \textit{Kepler} Transit 21 centered around time 610.4 Barycentric \textit{Kepler} Julian Date (BKJD, i.e. BJD - 254833).  This is the simplest spotted transit for the system, showing only one distinct spot feature during the transit.  The initial MCMC optimization run found a single large spot (centered around 610.6 BKJD) almost entirely in the path of the secondary, as seen in Figure~\ref{fig:T21} (bottom).  The model light curve shown in red (Figure~\ref{fig:T21}, top) results in a reduced $\chi^2 = 4.5$ relative to the \textit{Kepler} data.   We decided this fit was sufficient and no attempts with additional spots were necessary. With a best fit R$_{\rm spot}/$R$_*$ = 0.16$~\pm~$0.01, this spot is $\sim 70\%$ larger than the largest ever sunspot (relative radius of 11\% the radius of the Sun) \citep{newton1955}. 

Typically, the KOI-340 transits have more complex structures than shown with Transit 21. If the initial one- or two-spot run cannot match all the subtle features in the light curve, we add additional spots and apply the MCMC optimizer in 4000 step runs until the best fit \texttt{STSP} model matches the data well and has a reduced $\chi^2$ less than 10.  Depending on the complexity and number of spots, each transit typically takes 10 of these \texttt{STSP} runs to satisfy the convergence criterion, making the total steps taken approximately 40,000 steps.

\begin{figure}[h]
    \centering
    \begin{minipage}{0.48\textwidth}
        \centering
        \includegraphics[width=\textwidth]{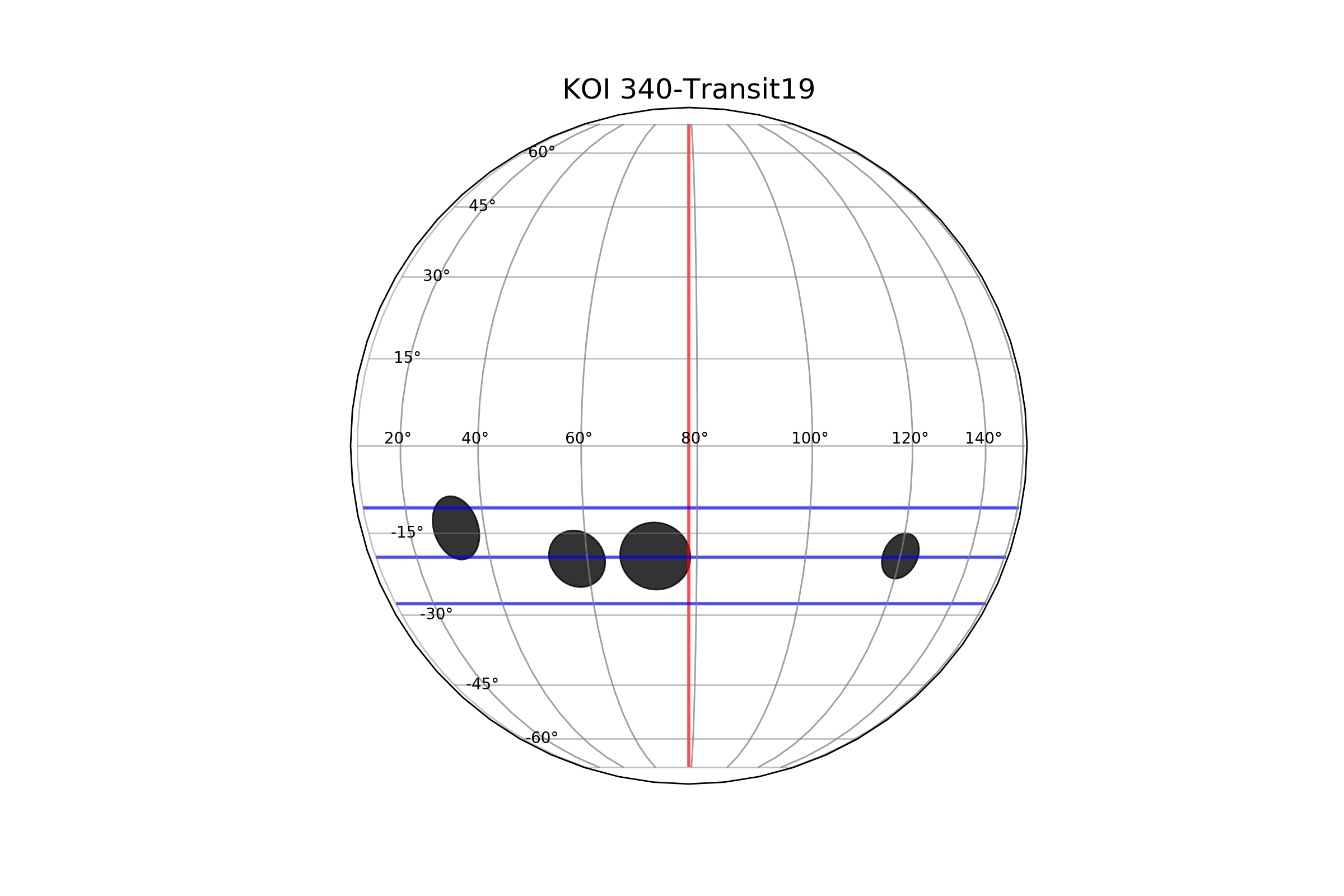} 
    \end{minipage}%
    \begin{minipage}{0.48\textwidth}
        \centering
        \includegraphics[width=\textwidth]{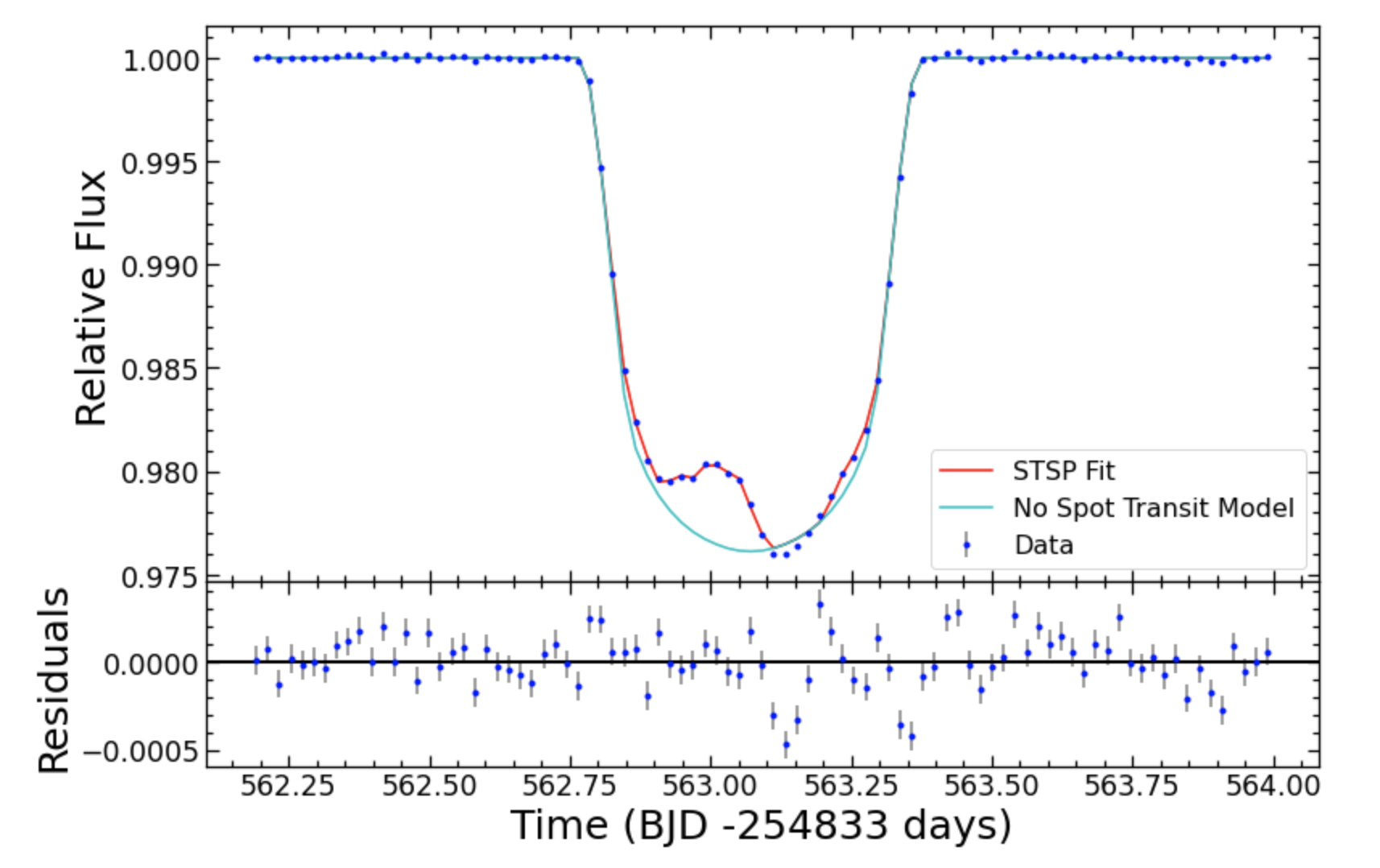} 
        \caption{Similar plot to Figure \ref{fig:T21} except for Transit 19.}
    \end{minipage}
    \label{fig:T19}
\end{figure}

Transit 19 is representative of most KOI-340 transits, with the final \texttt{STSP} modeling results shown in Figure \ref{fig:T19}. For this transit, we initially modeled it using initial conditions in \texttt{STSP} with two spots. However, we quickly found that two spots were not sufficient to fit the complex bump feature (centered around 563.00 BKJD) along with the bumps on ingress and egress (centered around 562.8 BKJD and 563.25 BKJD respectively). Once we increased the number of spots to four, we were able to fit the primary transit very well with a final reduced $\chi^2$ value of 2.2. In this case, the shorter, wider bump (at time 563.0 BKJD) is best fit with two smaller spot groups that are located right next to each other rather than one large spot group. This is in comparison to the sharper, larger bump seen in Transit 21 which was fit very well with only one spot. As seen in Figure \ref{fig:T19}, all of the spot groups modeled with \texttt{STSP} for this transit are well within the path of the companion, and all four of the spots have radii from R$_{\rm spot}/$R$_{*} = 0.07 - 0.10$ which are comparable to Solar maximum sunspots (see Section \ref{sec:radii} for more details on typical sunspot sizes) \citep{howard1984}. 

\begin{figure}[h]
    \centering
    \begin{minipage}{0.48\textwidth}
        \centering
        \includegraphics[width=\textwidth]{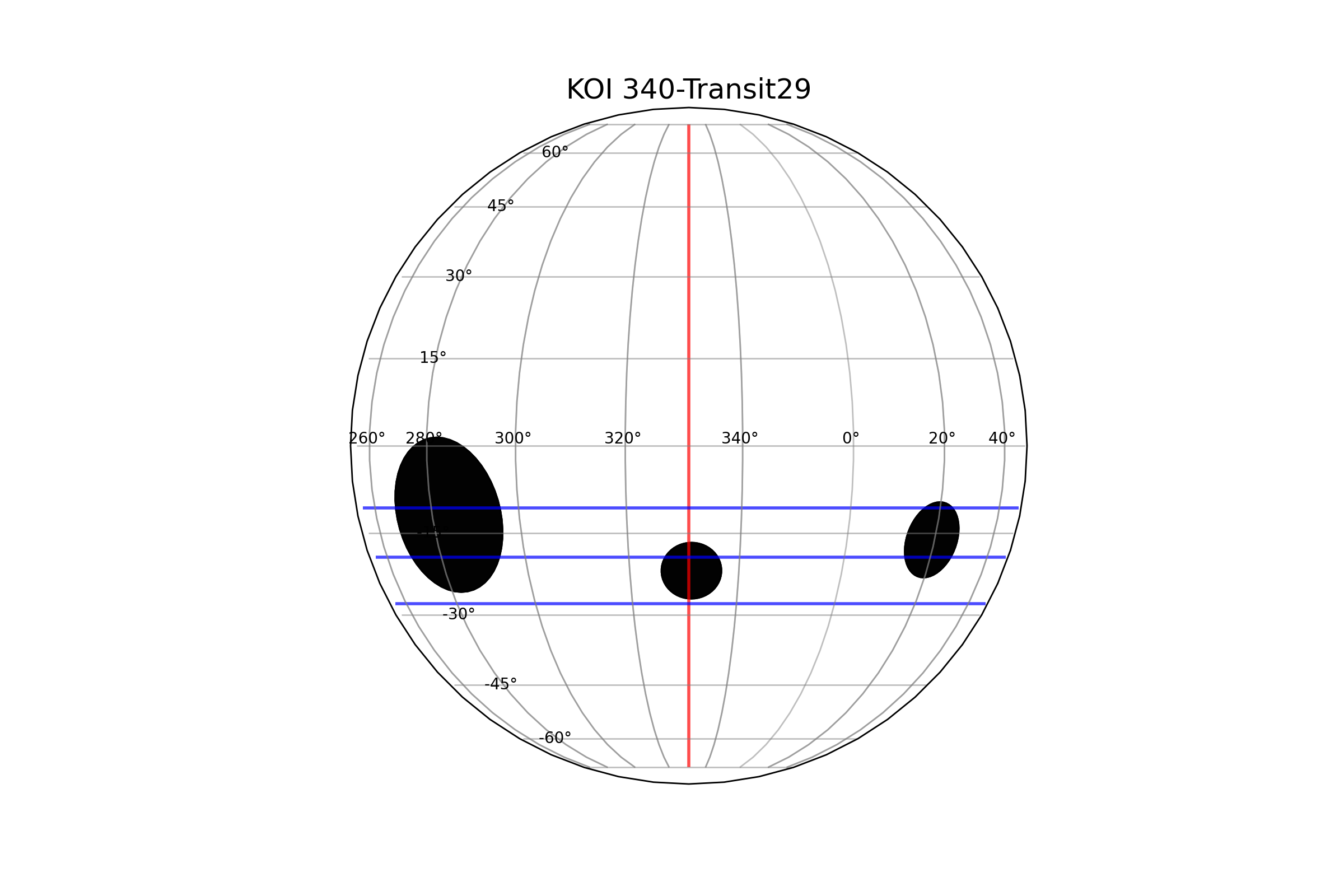} 
    \end{minipage}%
    \begin{minipage}{0.48\textwidth}
        \centering
        \includegraphics[width=\textwidth]{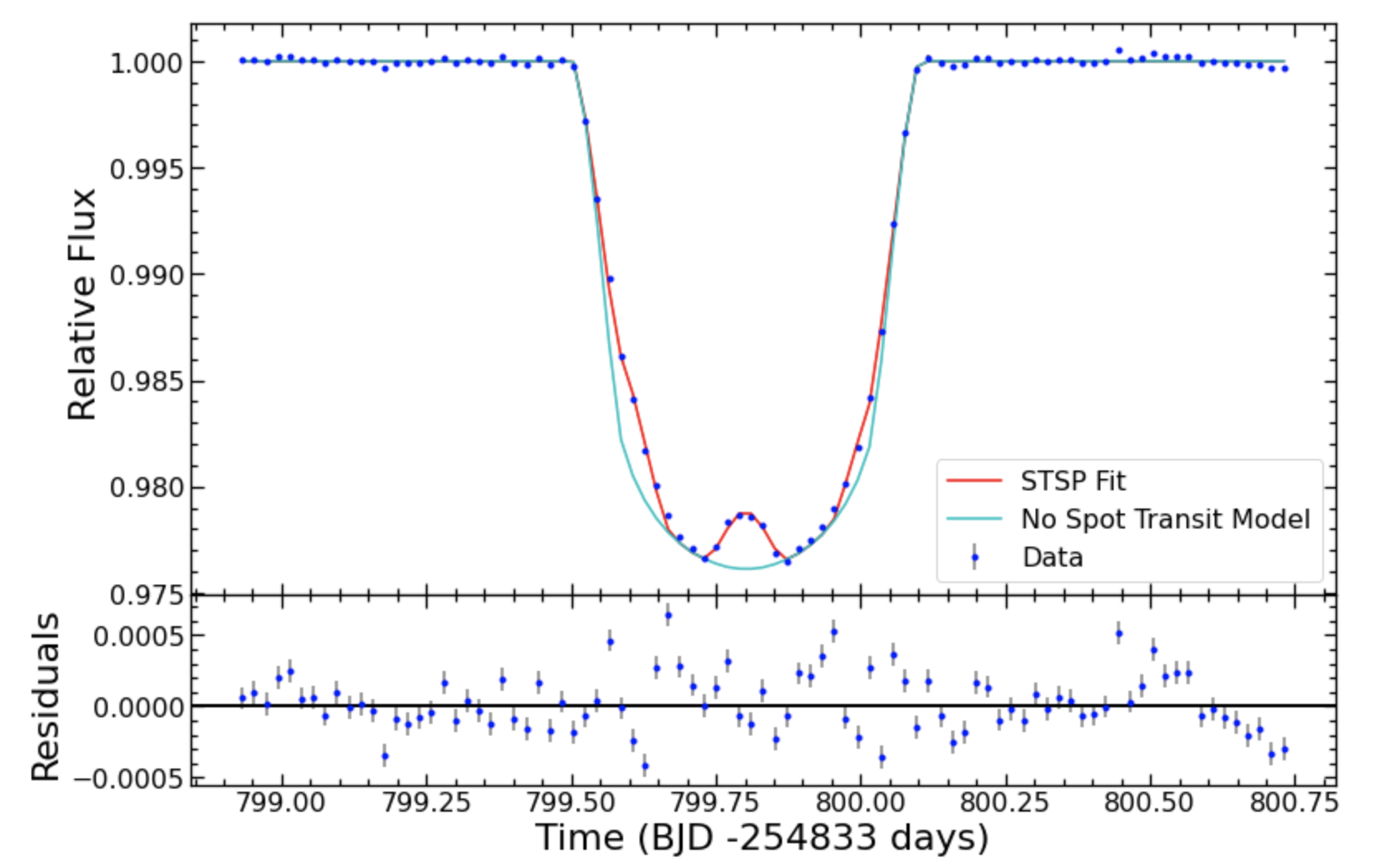} 
        \caption{Similar plot to both Figure \ref{fig:T21} and Figure \ref{fig:T19} except for Transit 29.}
    \end{minipage}
    \label{fig:T29}
\end{figure}

One of the key features in many KOI-340 transits is spot groups on the ingress and egress of the transit as shown in Transit 19. However, spots on the ingress and egress can lead to very different final spot parameters depending on the duration of the feature. Transit 29 showcases this phenomenon as seen in Figure \ref{fig:T29}. For the spot on the ingress of Transit 29 (centered around 799.6 BKJD), it extends over five cadences (150 minutes total for \textit{Kepler} long cadence data) and is very distinct from the no spot model. Thus, \texttt{STSP} finds the best fit spot to be both large in size and mostly in the path of the secondary (R$_{\rm spot}/$R$_*$ = 0.234). In comparison, the spot on the egress of the transit (centered around 800.0 BKJD) only lasts for three cadences (90 minutes) and is not as distinct from the no spot model in Figure \ref{fig:T29} (cyan line) leading to a smaller spot that is fully in the path of the secondary (R$_{\rm spot}/$R$_*$ = 0.115). Both of these spots are distinct from the small bump in the middle of the light curve centered around time 799.8 BKJD. Finally, for the small bump in the middle of the transit, \texttt{STSP} finds the best fit to be a smaller spot (R$_{\rm spot}/$R$_*$ = 0.089) fully in the middle of the secondary crossing path as expected. As all of the spots are distinct from each other, it is much easier to determine that there are only three spots in this transit, in contrast to the complex spot structure centered around 563.0 BKJD in Transit 19 (see Figure \ref{fig:T19}). The smaller bump in the middle of the transit is also a good contrast to the sharper, taller bump seen in Transit 21 (centered around 610.4 BKJD, see Figure \ref{fig:T21}), which gave rise to a much larger spot radius.

There is a known degeneracy between the central latitude and radius of a spot because at the precision and sampling of the \textit{Kepler} data most features that can be fit by a small spot that is fully in the path of the transit chord can be equally well fit by a larger radius spot that is barely grazing the transit \citep[see Figure 5]{morris2017}.  To mitigate this problem, we manually select the best fit results that favor smaller, in-transit spots as opposed to grazing spots, and we provide error bars for the relative radius of the spot, which accounts for these near-equivalent solutions.  The error bars are calculated from the MCMC output using the \texttt{corner.py} software which calculates the appropriate 2D Gaussian density rather than the regular 1D 1$\sigma$ error bars \citep{corner}.

\section{Results and Discussion}\label{sec:res_dis}

We used \texttt{STSP} as described in Section~\ref{sec:STSP} to model 36 tranists of KOI-340 that show evidence of spot occultations during a transit. Two of the 38 total transits show no signs of surface brightness variations and were not modeled with \texttt{STSP} with two additional transits showing very little sign of variation that were modeled using \texttt{STSP} giving a 89.4\% probability of strong starspot crossing features during a primary transit for KOI-340.  

The best fitting transit models for the 36 transits produce 122 total starspots in the path of the planet, each defined by its position on the surface of the star and its size relative to the stellar radius (latitude, longitude and R$_{\rm spot}$/R$_*$).  It is important to note that the spots identified on KOI-340 are likely to be starspot groups like the active regions on the Sun, rather than individual starspots, given their large sizes. 

\subsection{Spot longitudes and latitudes}

\begin{figure}[h]
\centering
\includegraphics[width=0.48\textwidth]{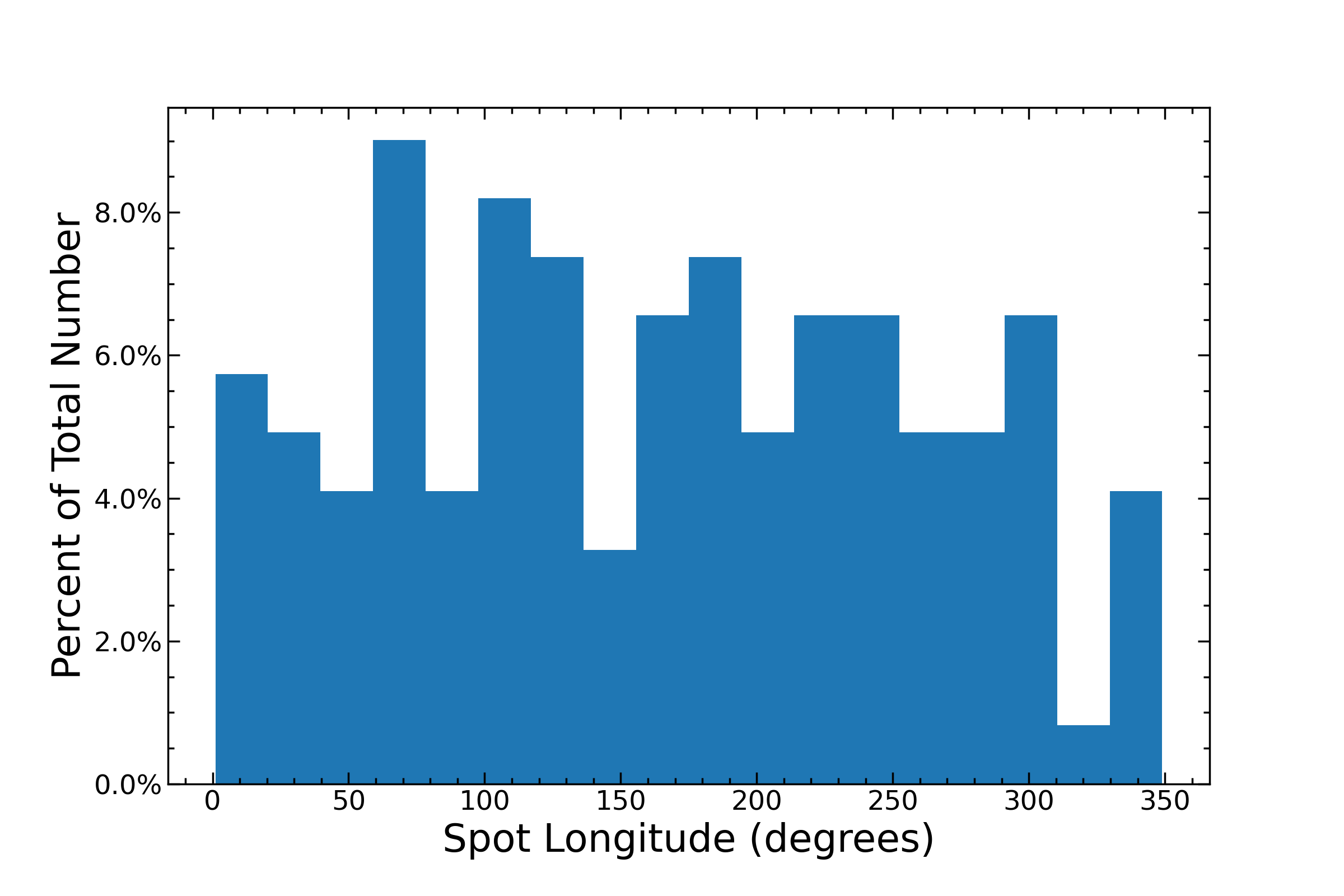}
\caption{Distribution of spot group central longitudes for KOI-340 is shown here in blue. This distribution shows no true peak meaning KOI-340 has spots at every longitude equally. As there are no longitudes that are highly favored, it is unlikely the M dwarf companion is inducing spots on the surface of KOI-340 that are large enough to be detected over the rotation induced spots. }
\label{fig:long_hist}
\end{figure}

\begin{figure}[h]
\centering
\includegraphics[width=0.48\textwidth]{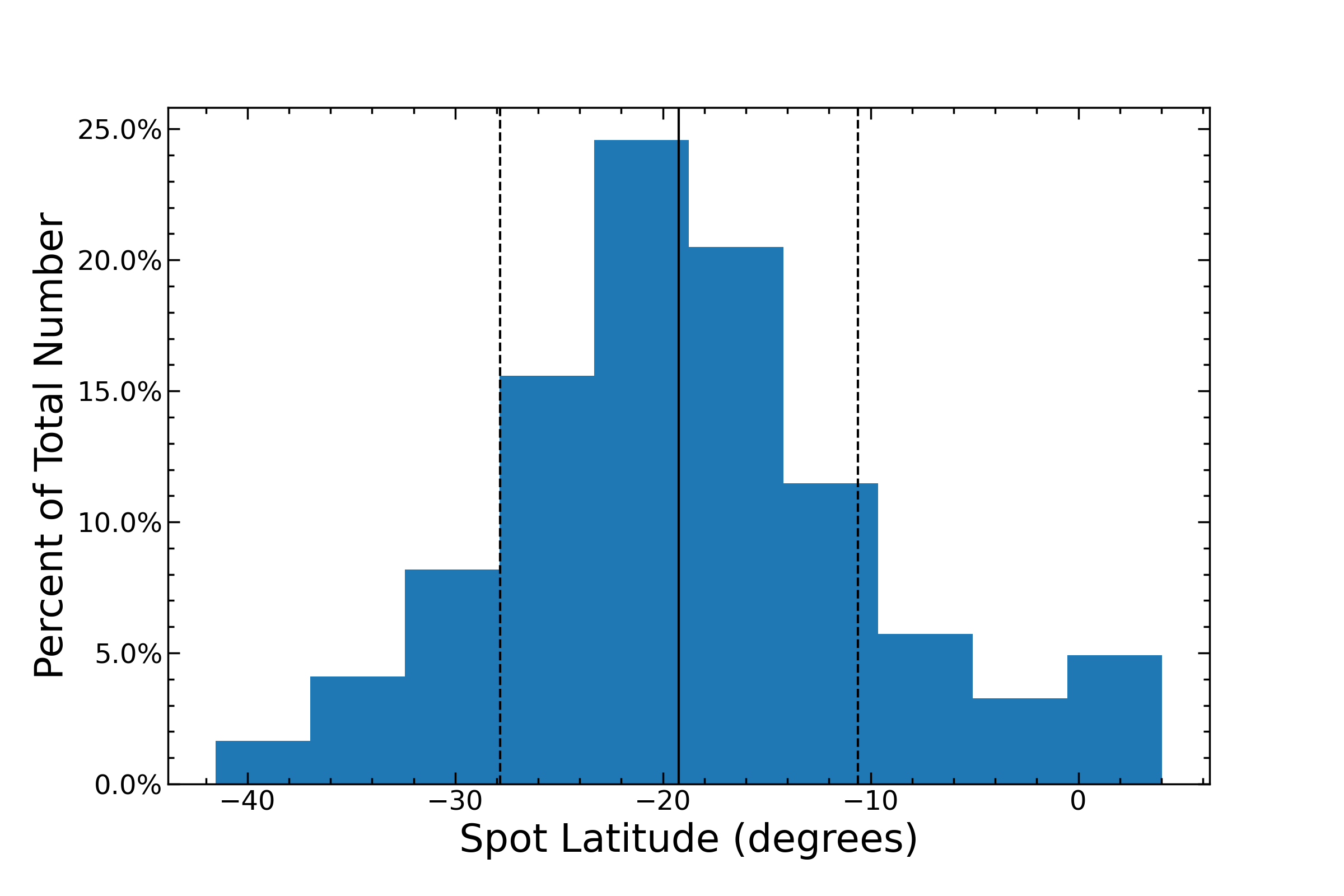}
\caption{Distribution of spot group central latitudes for KOI-340 is shown here in blue. The distribution is centered at $-19.3^\circ$, which coincides with where the M dwarf companion crosses (solid black line). The dashed black lines shows the full extent of the secondary crossing path. 24\% of the spots fall outside the dotted lines and are thus grazing spots.}
\label{fig:lat_hist}
\end{figure}

In Figures \ref{fig:long_hist} and \ref{fig:lat_hist}, we plot the distribution of spot longitudes and latitudes, respectively.  The longitude distribution reveals that there are spot groups occurring at every longitude, and spot occultations in nearly every transit, and thus there is no preferred longitude for spots detected at mid-latitudes in the path of the secondary. We also do not find any evidence of two preferred active longitudes that are 180$^{\circ}$ apart from each other as has been seen in multiple other types of active stars like RS CVn type stars \citep{berdyugina1998}, FK-Com type stars \citep{jetsu1993}, young, active Solar analogues \citep{berdyugina2002}, and even in the Sun \citep{berdyugina2003}. However, active longitude studies typically involve long term photometric observations which we do not have for KOI-340. In comparison, the histogram of latitudes of the spot groups (see Figure \ref{fig:lat_hist}) illustrates a more defined distribution, as expected because modeling in-transit spots is only sensitive to the portion of the host star that is covered by the path of the transiting object. The center of the transit path is shown in Figure \ref{fig:lat_hist} as a solid black line with the full extent of the transit path shown with dashed black lines.  24\% (29 of the 122 total spots) of spots have their central latitude outside of the path of the planet (i.e. they don't fall in the dashed line region).  We refer to these as grazing spots.   

\subsection{Spot Radii}\label{sec:radii}

\begin{figure}[h]
\centering
\includegraphics[width=0.48\textwidth]{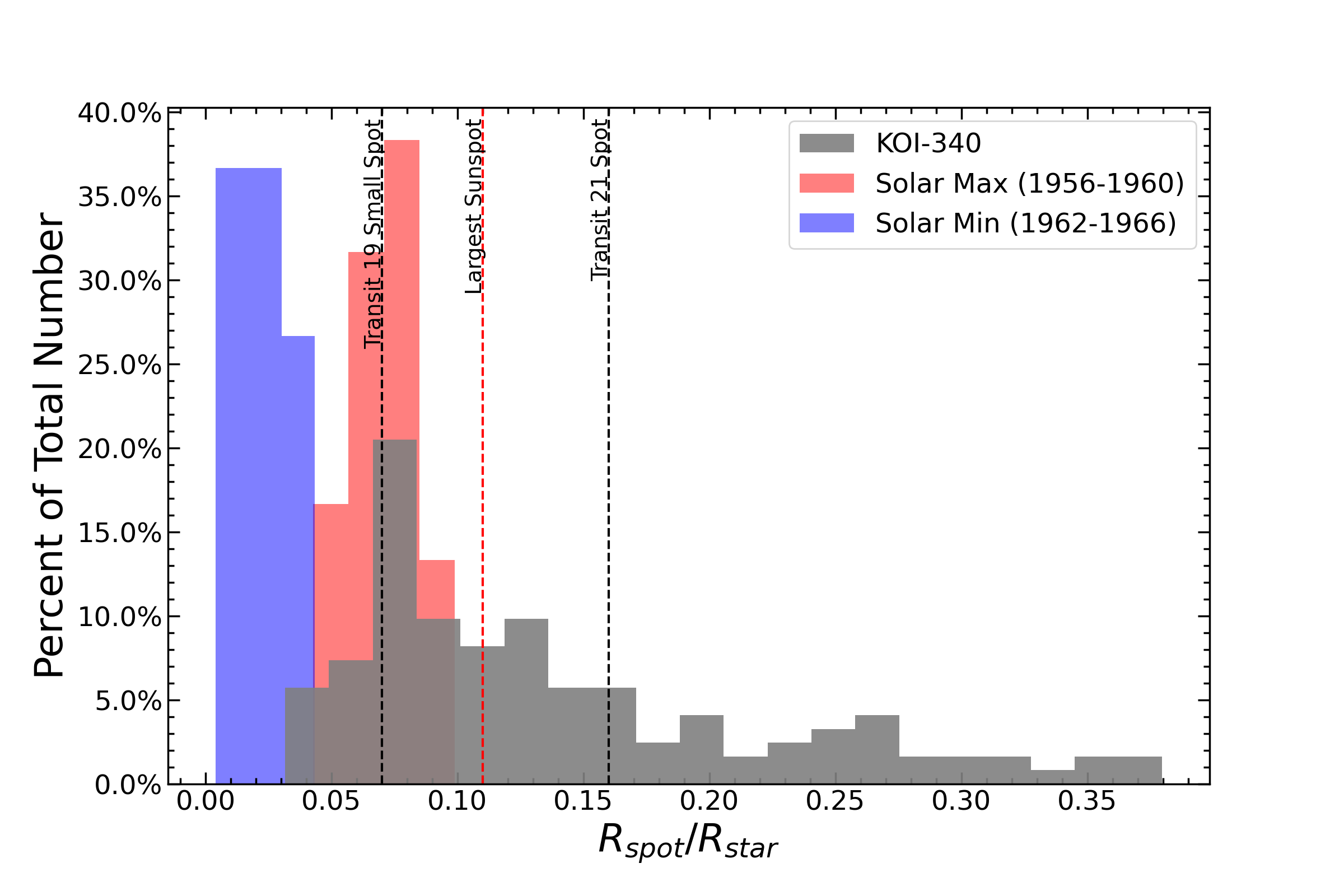}
\caption{KOI-340's spot radius distribution is shown in grey with spot radius distributions for typical Solar maximum and Solar minimum sunspots over same duration as the \textit{Kepler} mission (four years). Here the spot radius is given in relative radius, $R_{\rm spot}/R_{*}$ with a bin size width of 0.018 for all three distributions. For comparison, the black dotted vertical lines correspond to the smallest spot (centered around 563.25 BKJD) found in Transit 19  and the main spot (centered at 610.4 BKJD) found in Transit 21 (see Figures \ref{fig:T19} and \ref{fig:T21}), and the red dotted line corresponds to the largest sunspot ever found by \citet{newton1955} converted to relative radius as shown in \citet{morris2017}.}
\label{fig:rad_hist}
\end{figure}

We next compare our distribution of spot sizes for KOI-340 to a representative distribution of sunspots at Solar maximum and Solar minimum with the same baseline as the \textit{Kepler} mission of 4 years (Figure \ref{fig:rad_hist}). For the Solar maximum distribution, we used a subset of data from 1956-1960 shown in red, and for the Solar minimum distribution, we used a subset of data from 1962-1966 shown in blue \citep{howard1984}. These subsets were chosen because they are periods of time that are solely Solar maximum or minimum. 
 
The distribution of KOI-340's spot size peaks with a value around 7.5\% of the star's radius, which is similar to the peak of the Solar maximum distribution.  These spots are the most common sized spots on KOI-340 as well as the Sun during Solar maximum.  The smallest spot shown in Figure~\ref{fig:T19} during Transit~19 is an example of this typical spot.           
However, KOI-340 has many more larger spots than the Sun as evidenced by the tail of the distribution that extends out to a relative radius of $\sim 1/3$ the size of the star.  KOI-340's median spot radius is R$_{\rm spot}/$R$_*$ = $0.1144$ making half of the spots detected on KOI-340 larger than the largest ever sunspot (11\% of the Sun, \citealt{newton1955}).    
Despite the degeneracy between spot radius and latitude (described above), only 29 of the very largest spots in the 122 spot distribution are grazing spots that may appear artificially large.  Discarding all spots with centers not in the path of the transiting companion, the distribution still extends out to R$_{\rm spot}/$R$_*$ = $0.29$ which is 2.6 times the largest ever sunspot. This tail of larger spots could also be due to unresolved spot groups which causes the spot groups to appear larger.  

\subsection{Fractional Area}\label{sec:fraca}

In addition to spot radii, we also calculate a lower limit on the fraction of the star that is covered by spots at the time of each transit.  This quantity is important because it is immune to the spot-radius degeneracy, and it is necessary for transmission spectroscopy analyses when calculating the depth correction to the transit due to starspots.  Indeed, \citet{pont2007} showed that for even a total spotted area of 1\%, starspots would be the dominant source of uncertainty for a transmission spectrum of an extrasolar planet. \citet{rackham2018} showed that for the very active M-dwarf TRAPPIST-1 system, with spot covering fractions around 8\%, the resulting stellar contamination affects the transit depths 1-15 times more than planetary atmospheric features. However, observations of the system have yet to produce evidence for significant transit contamination by dark starspots \citep{Ducrot2018,Morris2018b,Gressier2022}, while bright regions have been proposed to explain the rotational modulation of the host \citep{Morris2018c, Wakeford2019}.

\begin{figure}[h]
\centering
\includegraphics[width=0.48\textwidth]{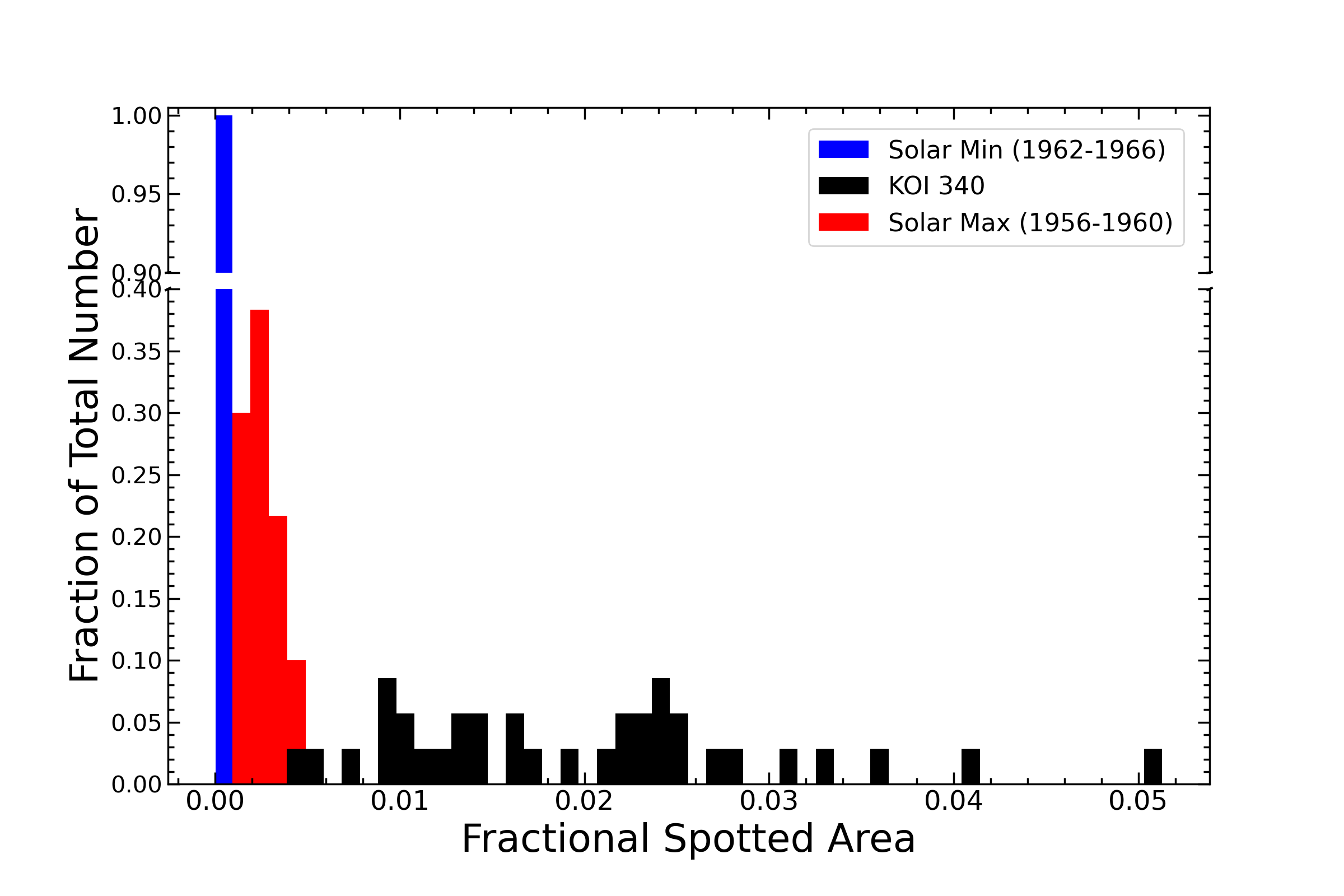}
\caption{Area of transit crossing chord that is spotted assuming there are no spots anywhere else on the front hemisphere of KOI-340 is shown as the black distribution. This distribution is then the minimum fractional spotted area for the front hemisphere of KOI-340 compared to the Solar maximum fractional spotted area (red distribution) and Solar minimum fractional spotted area (blue distribution). The Solar maximum and minimum distributions are the same time frames as shown in Figure \ref{fig:rad_hist}. All three distributions have bin sizes of 0.001 (or 0.1\%) fractional spotted area. The y-axis has been broken from 0.4 until 0.9 as the Solar minimum distribution has much smaller fractional spotted areas so all of the values are in one bin.}
\label{fig:frac_hist}
\end{figure}

We calculate the area of each circular spot that falls in the path of the companion and sum over all spots.  We then divide the total spotted area in the transit chord by the total area of the hemisphere of the star ($2\pi R_*^2$).   This quantity represents the minimum fractional area of the stellar surface covered by spots because it assumes no other spots exist on the front face of the star.  Figure~\ref{fig:frac_hist}, compares the lower limits derived for all transits of KOI-340 to the monthly average values for the Solar minimum (blue) and Solar maximum (red) distributions (as defined in above).  We use the monthly average because the orbital period of KOI-340 ($\sim 23$d) allows us to take a snapshot of the stellar surface approximately once a month over the 4-year duration of the \textit{Kepler} survey.      

KOI-340’s fractional spotted area ranges from 0.4\% to 5\% of the stellar surface, and is almost always greater than the Sun’s.  
All but one snapshot show a minimum spot covering fraction that is greater than the spot covering fractions over the whole Sun at any point during its cycle.  
The mean value for the minimum spot covering fraction on KOI-340 is 0.0198 Hemispheres (Hems), which is $\sim$10 times greater than the largest Solar fractional spotted area ever recorded \citep{cox2000} and large enough to create significant uncertainty in any transmission spectra of planets orbiting stars like KOI-340.  


\begin{figure*}[t]
\centering
\includegraphics[width=\textwidth]{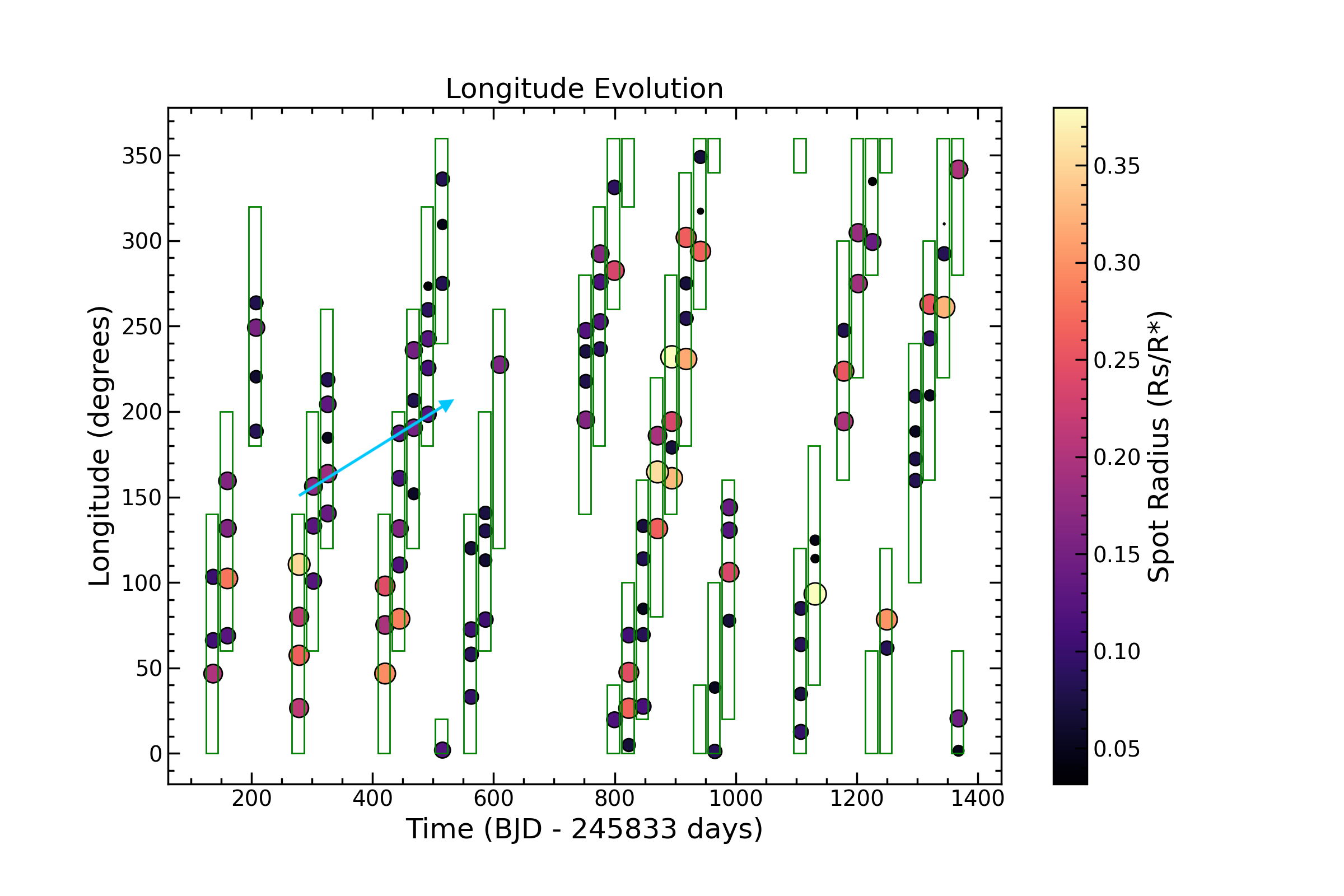}
\caption{Plot of longitude versus time of all 122 spots modeled for KOI-340 with the markers being sized with respect to the relative radius of the spot group and colored with relative radius as well. The green boxes surrounding each transit correspond to the longitude coverage for each transit. When the green boxes overlap with the next transit, we looked for signs of the spots surviving to the next transit in order to search for signs of differential rotation. An example of this is shown with the overplotted cyan arrow which encompasses two such instances of possible signs of spots moving in longitude over time.}
\label{fig:long_evol}
\end{figure*}

In Figure \ref{fig:long_evol}, we plot the longitude of every spot for a given transit versus the midpoint time of that transit. We have formatted the size of the marker to correspond to the radius of the spot group and colored the points according to relative spot radius.  The green boxes surrounding each transit correspond to the total longitude coverage for each transit.
In seven cases there are enough consecutive transits that we can combine every other transit in groups of three to create a complete 360 degree view of KOI-340 in longitude space (the transits centered around 400, 450, and 500 BKJD for example). By using these seven total instances, we can compute a lower limit for the fractional area of the entire star in the transit crossing region. In doing so, we get a range of fractional areas for the entire star in the transit crossing path to be 2.1-4.3\%. If we assume that KOI-340 were to act similarly to the Sun, then we might expect KOI-340 to have a matching band in the Northern hemisphere of the star that is similarly active to the region of the transit crossing which would give a total fractional spotted area for the entire star of 4.2-8.6\%. If we simply scale up the minimum total fractional spotted area for the entire star of 2.1\% in the transit crossing path to be the same across the entire star, we get a value of $\sim$14\% for an estimate of the total fractional spotted area of KOI-340.

\subsection{Longitude Evolution}\label{sec:long}

Figure \ref{fig:long_evol} is also used to determine how the spots evolve over time. If two spot features are detected at similar longitudes in consecutive transits, this provides evidence that those features are caused by the same spot that has survived for more than the orbital period of the companion (23 days). Of the 122 total spots for KOI-340, there are 54 possible spots that could be seen in the next transit, and in 30 of those instances, a second spot feature is detected at a similar longitude in the following transit. Thus, there is a 55\% probability that the spots persist to the consecutive transit meaning there is a 55\% chance of the spots living longer than 23 days.  In comparison, sunspots typically live for around a day but can live up to months depending on the size with larger sunspots lasting for longer amounts of time \citep{solanki2003}.

Furthermore, we use Figure \ref{fig:long_evol} to estimate the differential rotation at a latitude of $-19.3^\circ$ on KOI-340 by quantifying the progression in longitude of the 30 spots that are observed in consecutive transits.  The majority of these spots progress forward in longitude, suggesting that the rotational motion of the latitude where the companion crosses is moving faster than the average rotation period defined by spots at all latitudes \citep{gj12432015}. Estimating the differential rotation from the slope of the cyan line provided in Figure \ref{fig:long_evol}, we get a value of $\Delta\Omega \sim 0.004 \pm 0.001$ rad day$^{-1}$ which is an order of magnitude lower than the Sun's differential rotation value of $\Delta\Omega = 0.055$ rad day$^{-1}$ \citep{bergyugina2005}. Presumably, the rotation period of KOI-340 measured from the periodogram is also generated by some of the same spots as we use here to infer the differential rotation implying the rotation rates should be similar between the two measurements. This is consistent with the small value we measure here for the differential rotation. Thus, this estimate is only a weak lower limit. Also, the long orbital period of 23 days for KOI-340's companion provide a sampling time that is not fine enough for a definitive measurement furthering our assumption this is a weak lower limit. With other stars like GJ 1243 in well-sampled \textit{Kepler} data and with careful \texttt{STSP} modeling of the in-transit variability, there are signs of progression in longitude of the spots leading to an indication of differential rotation on the surface of GJ 1243 \citep{gj12432015}. 

Finally, we also investigated if the radius of the spot changed at all from one transit to the next for the 30 spots that survive consecutive transits and found that the spots do not grow or shrink in any significant way. Unlike sunspots which decay more rapidly over time, the spots on KOI-340 appear to not change in size significantly if they survive for one complete orbital period (23 days). 

\subsection{Modeling of Plage}
\begin{figure}[h]
    \centering
    \begin{minipage}{0.48\textwidth}
        \centering
        \includegraphics[width=\textwidth]{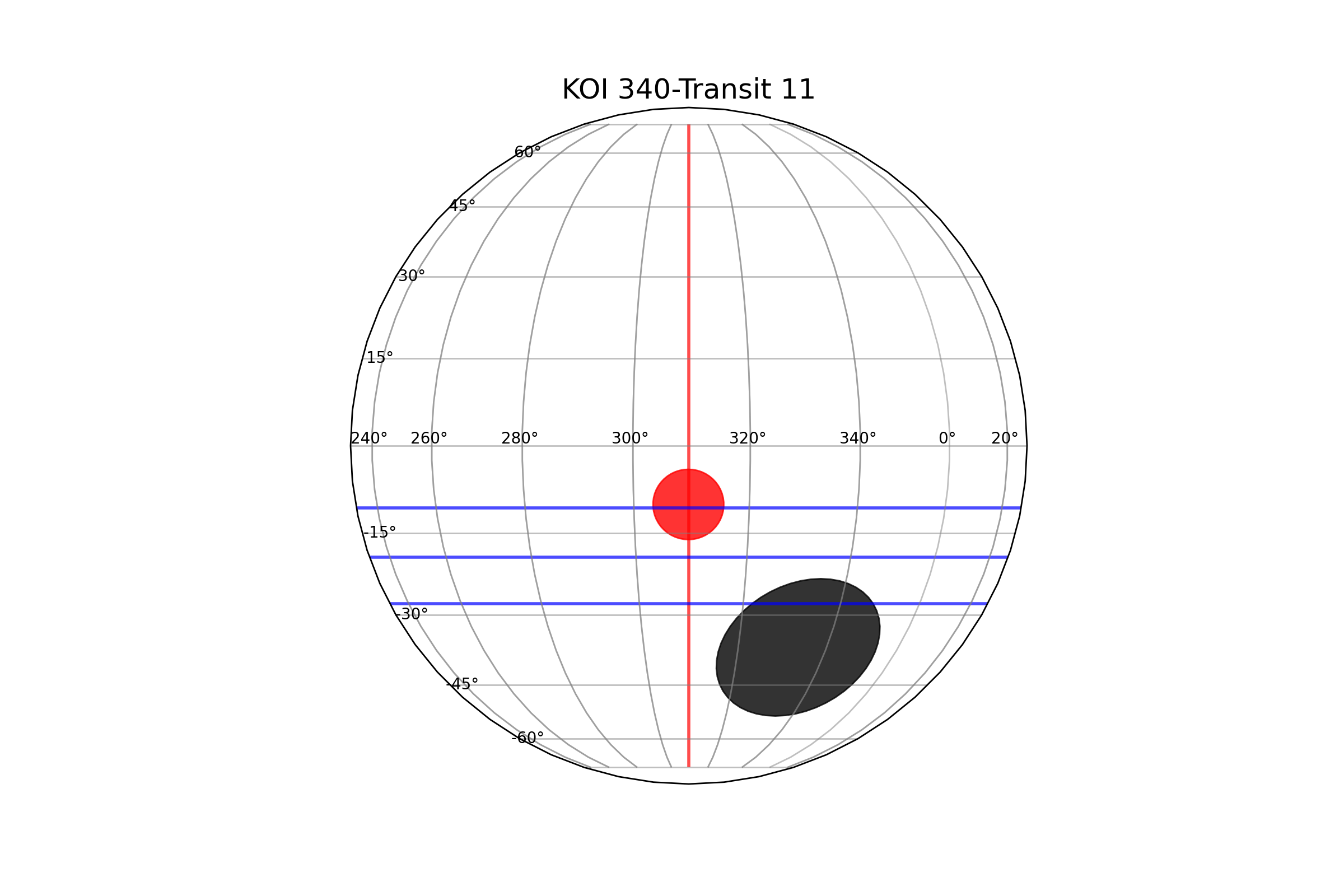} 
    \end{minipage}%
    \begin{minipage}{0.48\textwidth}
        \centering
        \includegraphics[width=\textwidth]{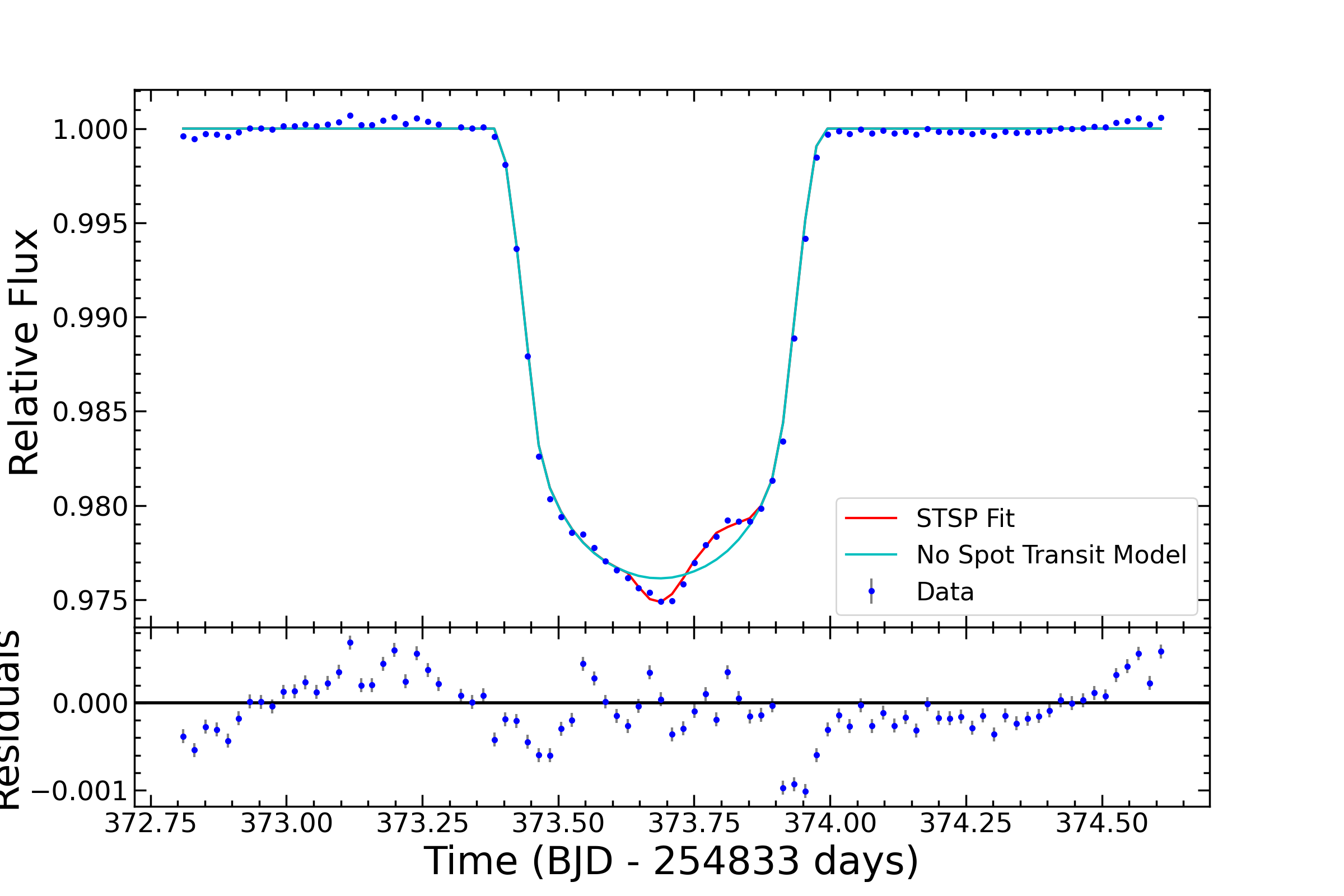} 
        \caption{Top: Plot of the surface of KOI-340 with the final spot groups shown as black filled circles along with the red line denoting the longitude of the star at mid-transit and with blue lines denoting the full extent of the transit path for the companion. Red circle denotes the bright spot modeled in this transit. Bottom: Light curve for final \texttt{STSP} fit (red line) along with the no spot model for KOI-340 (cyan line) for Transit 11. The residuals (model - data) are shown below the light curve with blue points.}
    \end{minipage}
    \label{fig:T11}
\end{figure}

In one instance during a primary transit, we were able to model a bright stellar surface feature for KOI-340. During this transit, there was a dip significantly below the expected no-spot transit model as seen in Figure \ref{fig:T11} at around time 373.7 BKJD which indicates the presence of a plage rather than a darker spot as seen in the other transits. We implemented a new feature in \texttt{STSP} to allow for two contrasts to be specified during a modeling run. For the contrast of the bright spot, we chose to use a value of 1.3, that is 30\% brighter than the photosphere, as for the dark spots we used a value of c = 0.3 as described in \citet{morris2017}. The modeling of this transit was not unique in any other way, so our best fit model was still required to meet our convergence criteria mentioned in Section \ref{sec:STSP}. The best fit solution for the plage is a feature with $R_{\rm spot}/R_{*} = 0.10$. This plage seems to be closely followed by a larger, grazing spot, and it has been suggested that spots and plages are co-located on active G and K stars \citep{morris2018}.

\subsection{Out-of-transit Starspots}\label{sec:Simon}


Given that the secondary companion crosses the primary at a latitude of $-19.3^\circ$, the primary transit models that we have modeled and described in Section \ref{sec:STSP} only give us information about the surface features in the latitude range of $-27.8^\circ$ to $-10.6^\circ$ which encompasses the secondary's coverage on the host star. In order to study the surface features on the rest of the star, we model both the in- and out-of-transit light curve for one rotation period of KOI-340 centered on each primary transit. 
We model this entire light curve in \texttt{STSP} by fixing the previously determined in-transit spots and adding on additional spots until the brightness variations for the out-of-transit data are well modeled. As there is no constraint on the latitude of the added spots, the best fit spot positions and sizes from the out-of-transit variability provide degenerate results. 

Thus, we choose all chains with $\chi^2$ values within 20\% of the global minimum $\chi^2$ to be acceptable solutions. We then calculate the fractional spotted area of KOI-340 for all of the acceptable chains. Finally, we average the fractional spotted area across all acceptable chains to find the best fit fractional spotted area for one full rotation of KOI-340. For this procedure, we start with adding one additional spot and calculating the average fractional spotted area. However, the added spot needs to be large in size in order to match the variability seen in the out-of-transit data, so we repeat the same procedure adding additional spots until the fractional spotted area stops decreasing. Once the fractional spotted area appears to be constant even with an additional spot, we consider the out-of-transit modeling to be complete for that rotation. 

\begin{figure}[h]
\centering
\includegraphics[width=0.48\textwidth]{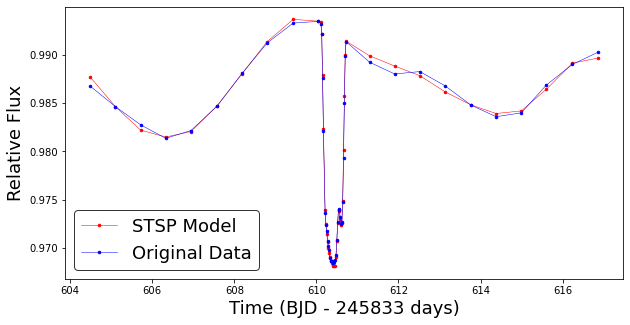}
\caption{Full light curve encompassing one full rotation period (12.96 days) for simplest primary transit model (Transit 21). The original \textit{Kepler} data is shown in blue with the \texttt{STSP} model shown in red. Three spots were needed (at minimum) to model the full light curve in addition to the one fixed in-transit spot.}
\label{fig:full_LC}
\end{figure}

\begin{figure}[h]
\centering
\includegraphics[width=0.48\textwidth]{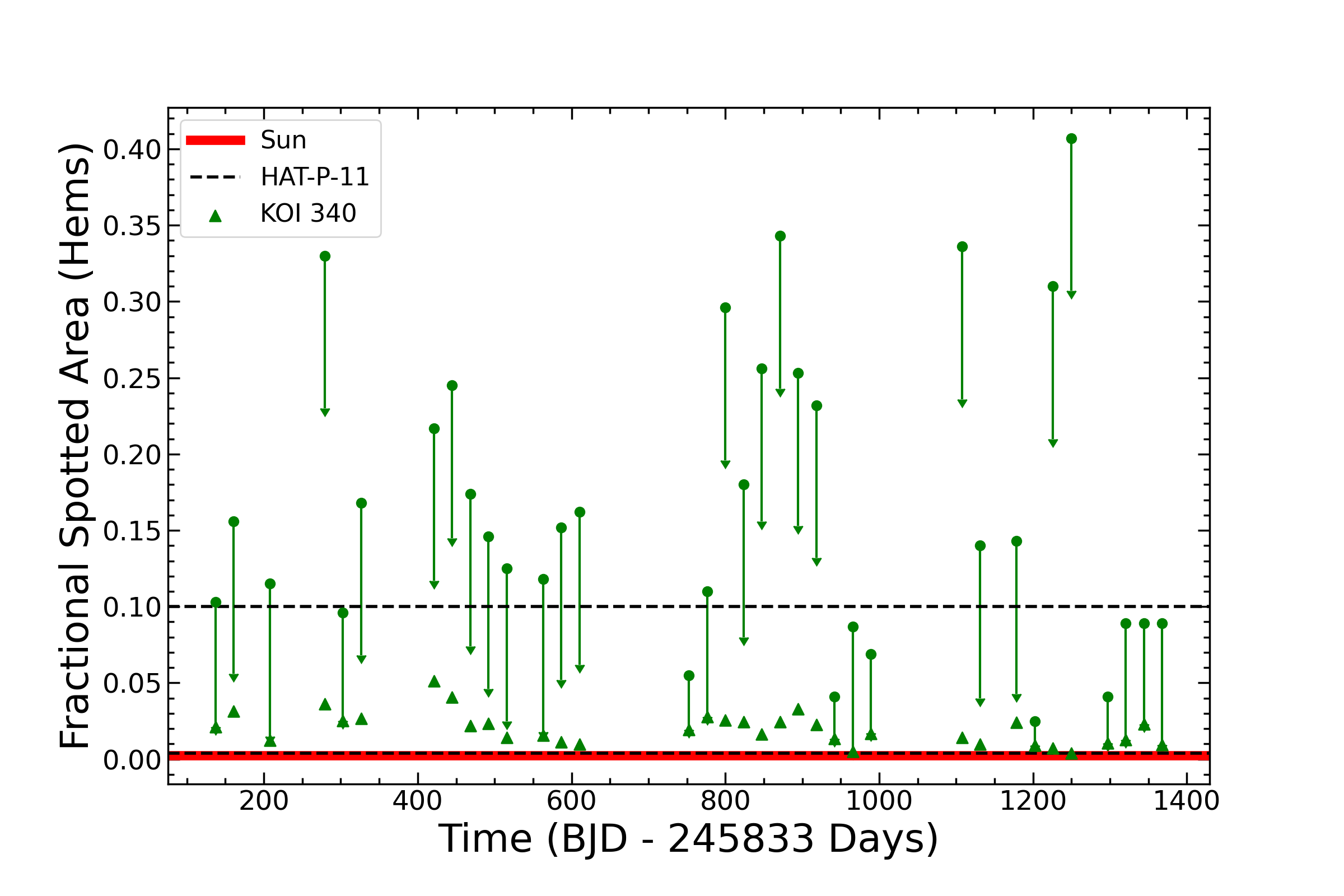}
\caption{Fractional spotted area plotted versus midpoint time of primary transits in BKJD. KOI-340 is shown using green symbols with the Sun's total fractional spotted area coverage given as a solid red bar \citep{howard1984} and HAT-P-11's range of fractional spotted areas given as two black dashed lines \citep{morris2017}. The green upper triangles denote the minimum fractional spotted area for that transit found using the in-transit spots, and the green points are positioned at the total fractional spotted area found using the full out-of-transit light curve modeled using the minimum number of additional spots as described in Section \ref{sec:Simon}.}
\label{fig:frac_a_vs_time}
\end{figure}

Currently, we have modeled all of the full rotations of KOI-340 with \texttt{STSP} with the minimum number of additional spots needed to match the full out-of-transit light curve for that rotation. An example of this full light curve modeling with the in-transit spots held fixed and the minimum number of spots added is shown in Figure \ref{fig:full_LC} with this full rotation of the star being centered around what we call Transit 21 (see \ref{sec:STSP}). Using the total fractional spotted area of the entire star for each full light curve (of which there are 36), we can estimate the possible upper limits of the fractional spotted area for KOI-340. This is shown in Figure \ref{fig:frac_a_vs_time} where we have plotted the fractional spotted area versus the midpoint time of the transit (i.e. the center of the full light curve) with the green upper arrows indicating the minimum spotted area found using only the in-transit spots and the green points with approximate error bars showing the total fractional spotted area of the star using the full light curve with the minimum number of spots added. In Figure \ref{fig:frac_a_vs_time}, the Sun's fractional spotted area is plotted as a solid red bar \citep{howard1984} along with HAT-P-11's range of possible fractional spotted area plotted as black dashed lines \citep{morris2017}. 

However, with adding only the minimum number of additional spots to the light curve, the best fit spots end up being large in size in order to fit the large out-of-transit variability as seen for example in Figure \ref{fig:full_LC}.

\begin{figure}[h]
\centering
\includegraphics[width=0.48\textwidth]{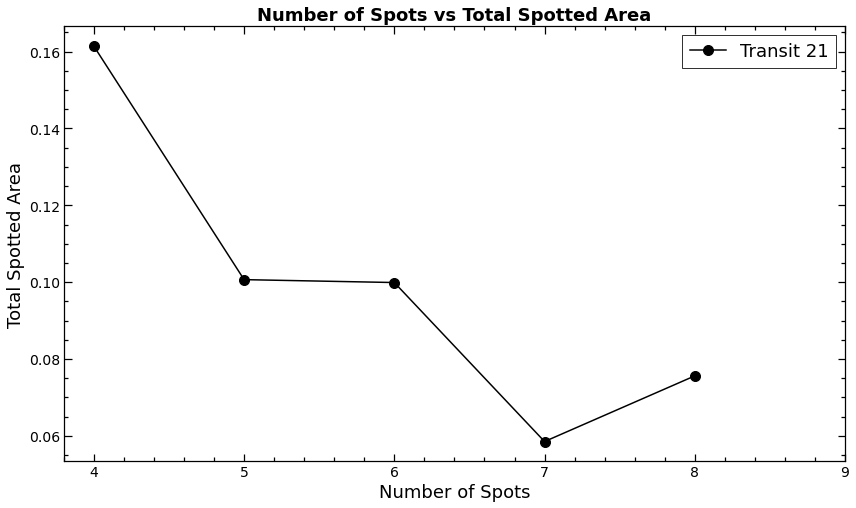}
\caption{Total spotted area for Transit 21's full out-of-transit light curve plotted versus total number of spots needed to fit the data. The in-transit spots are fixed at 1, so in total 7 additional spots were needed until the fractional spotted area started to level out. The minimum number of spots needed to fit the data was 3, so we added more spots on top of those 4 until the spotted area starts to level out meaning we are approaching the minimum total spotted area for the entire star.}
\label{fig:frac_a_t19}
\end{figure}

In order to determine a more accurate total fractional spotted area, we add additional spots and run \texttt{STSP} until we find a new best fit model. Then, we recalculate the fractional spotted area for KOI-340. In doing so, we can find the minimum fractional spotted area that still fits the full light curve which would then replace the current values shown as green points in Figure \ref{fig:frac_a_vs_time}. We have done this successfully so far for the simplest primary transit (Transit 21) as the total number of spots needed to fit the full light curve ended up only being 8. The upper limit error bars attached to the green points represent the $\sim$10\% spread found when we add additional spots to the model for this transit. Figure \ref{fig:frac_a_t19} shows the total spotted area versus the total number of spots, and when you add additional spots, the total spotted area decreases and then levels out indicating that we are approaching the minimum total spotted area. With every additional spot that \texttt{STSP} needs to model the complexity of the problem increases by a factor of 3 so the modeling runs end up taking 24-48 hours on average to complete rather than 8 hours for the regular in-transit modeling. Since the \texttt{STSP} modeling runs take days to complete, we leave the completion of the other 35 full light curves to future work.

\section{Conclusions: KOI-340 in larger context}\label{sec:conc}

KOI-340 is a G subgiant star with $\teff = \sim5600~K, \mstar = 1.21^{+0.04}_{-0.03}~\msun, \rstar = 1.89 \pm 0.05~\rsun$ with a rotation period of 12.96 $\pm$ 0.97 days and an M-dwarf secondary ($\mstar = 0.214 \pm 0.006~\msun$). KOI-340 has a cool temperature for its radius, and while there is no direct mass measurement for this star, \citet{brewer2018} estimates the mass from the temperature and radius to be around $1.2~\msun$ indicating that KOI-340 is likely in the process of evolving off the main sequence. KOI-340 also has a relatively slow rotation period for its estimated mass ($\sim 13$~days compared to $\sim 5$~days for main sequence stars \citep{mcquillan2014}) which could be due to angular momentum conservation as the star is getting larger. From the \texttt{STSP} modeling, we have determined that the radii of the in-transit spot groups can be much larger than typical Solar-maximum spots which might not be expected for a star of this mass and rotation rate. 

\begin{figure}[h]
\centering
\includegraphics[width=0.48\textwidth]{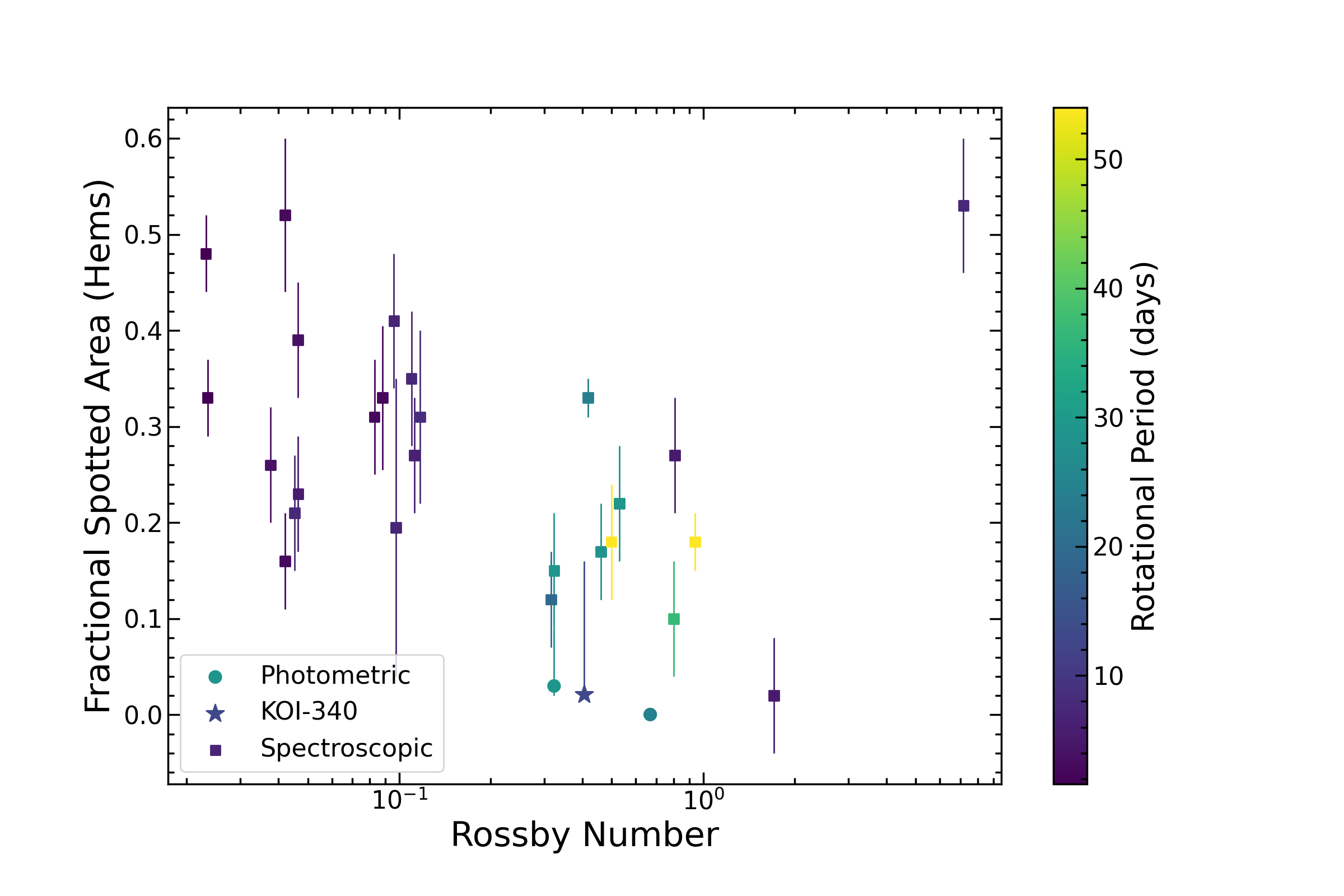}
\caption{Fractional spotted area of stars from \citet{oneal1998,oneal2001,oneal2004,morris2017,morris2019,howard1984} versus their Rossby number. The points are colored by their rotational period in days. Rossby numbers were calculated using their convective turnover time in days derived from their (B-V) color as done in \citet{mittag2018}. Stars that have their fractional area derived from spectroscopic methods are plotted using square symbols with the other photometric methods denoted by a circle symbol. KOI-340 is denoted with a star symbol around 0.4 in Rossby number with a dark blue color. The Sun is the blue point around 0.65 in Rossby number with no clear error bar, and HAT-P-11 is shown as both a light blue circle and square around 0.3 in Rossby number as it has both photometric and spectroscopic fractional spotted area measurements \citep{morris2017,morris2019}.}
\label{fig:comp_rot}
\end{figure}

In Figure \ref{fig:comp_rot}, we have plotted a variety of stars with estimated fractional spotted areas derived from other methods using spectroscopic features \citep{oneal1998,oneal2001,oneal2004}, other stars that have been modeled with \texttt{STSP} \citep{morris2017}, and the Sun \citep{howard1984} versus their Rossby number \citep{kim1996}. Their Rossby number was calculated using the equations in \citet{mittag2018} which gives an empirical equation for a star's convective turnover time as a function of its color. The points and error bars are colored by their rotational period, and the square symbols denote stars whose filling factor was determined by spectroscopic methods while the circle symbol denotes stars whose fractional spotted area was found through photometric methods like \texttt{STSP}. Figure \ref{fig:comp_rot} shows that typically the lower the Rossby number the higher the fractional spotted area. This trend makes physical sense as a lower Rossby number indicates a system with fast rotation and/or larger convection zones, which are therefore more likely to be active systems. Therefore, the large spots on KOI-340 compared to the Sun could be due to its faster rotation and increasing convection zone depth as it evolves off the main sequence.

It is also important to note that for KOI-340 we have chosen to use the mean minimum fractional spotted area (2.1\%; see Section \ref{sec:fraca}) even though some of the individual transits can have much higher estimated fractional spotted areas assuming certain conditions. We have chosen to use this conservative lower limit as we are only sensitive to the transit crossing path using our in-transit \texttt{STSP} models though once we have fully investigated the out-of-transit variability for KOI-340 as well we will be more sensitive to the entire star's spotted area. If the entire star was spotted at a similar rate, that gives a value of 14\% shown as the extent of the upper error bar for KOI-340. A more accurate measurement of the fractional spotted area will come from fully understanding the out-of-transit variability of KOI-340.

In summary, KOI-340 is an eclipsing binary system consisting of a G subgiant with an M dwarf companion that has starspots two times larger than the majority of sunspots on average. We used data from the \textit{Kepler} spacecraft to model both starspots and one plage using the starspot modeling program \texttt{STSP} which measures the position and radius of the surface features. We modeled 36 \textit{Kepler} transits and found the minimum fractional spotted area of KOI-340 is $2\substack{+12\\ -2}~\%$ while the spotted area of the Sun is at most 0.2\%. The starspots on KOI-340 were found to be present at every longitude with possible signs of differential rotation seen in the evolution of spots along with a 55\% of spots longer than 23 days. Thus, KOI-340 is a G subgiant star with an M dwarf companion that has considerable stellar activity covering $2\substack{+12\\-2}\%$ of the primary star at a minimum. Future work includes fully understanding and modeling the out-of-transit variability of KOI-340 which will provide constraints on the total spot coverage of KOI-340. 

\section{Acknowledgements}
We thank our referee for providing us very constructive feedback that helped us to improve the clarity and content of this paper. We acknowledge support from NSF grant AST-1907622. This research has made use of the NASA Exoplanet Archive, which is operated by the California Institute of Technology, under contract with the National Aeronautics and Space Administration under the Exoplanet Exploration Program. This paper includes data collected by the \textit{Kepler} mission and obtained from the MAST data archive at the Space Telescope Science Institute (STScI). Funding for the \textit{Kepler} mission is provided by the NASA Science Mission Directorate. STScI is operated by the Association of Universities for Research in Astronomy, Inc., under NASA contract NAS 5–26555.

\newpage
\bibliography{references}

\begin{thebibliography}{}
\expandafter\ifx\csname natexlab\endcsname\relax\def\natexlab#1{#1}\fi
\providecommand{\url}[1]{\href{#1}{#1}}

\bibitem[{{Aceituno} {et~al.}(2013){Aceituno}, {S{\'a}nchez}, {Grupp}, {Lillo},
  {Hern{\'a}n-Obispo}, {Benitez}, {Montoya}, {Thiele}, {Pedraz}, {Barrado},
  {Dreizler}, \& {Bean}}]{aceituno2013}
{Aceituno}, J., {S{\'a}nchez}, S.~F., {Grupp}, F., {et~al.} 2013, \aap, 552,
  A31

\bibitem[{{Barclay} {et~al.}(2021){Barclay}, {Kostov}, {Col{\'o}n}, {Quintana},
  {Schlieder}, {Louie}, {Gilbert}, \& {Mullally}}]{barclay2021}
{Barclay}, T., {Kostov}, V.~B., {Col{\'o}n}, K.~D., {et~al.} 2021, \aj, 162,
  300

\bibitem[{{Benneke} {et~al.}(2019){Benneke}, {Wong}, {Piaulet}, {Knutson},
  {Lothringer}, {Morley}, {Crossfield}, {Gao}, {Greene}, {Dressing},
  {Dragomir}, {Howard}, {McCullough}, {Kempton}, {Fortney}, \&
  {Fraine}}]{benneke2019}
{Benneke}, B., {Wong}, I., {Piaulet}, C., {et~al.} 2019, \apjl, 887, L14

\bibitem[{{Berdyugina}(2005)}]{bergyugina2005}
{Berdyugina}, S.~V. 2005, Living Reviews in Solar Physics, 2, 8

\bibitem[{{Berdyugina} {et~al.}(1998){Berdyugina}, {Berdyugin}, {Ilyin}, \&
  {Tuominen}}]{berdyugina1998}
{Berdyugina}, S.~V., {Berdyugin}, A.~V., {Ilyin}, I., \& {Tuominen}, I. 1998,
  \aap, 340, 437

\bibitem[{{Berdyugina} {et~al.}(2002){Berdyugina}, {Pelt}, \&
  {Tuominen}}]{berdyugina2002}
{Berdyugina}, S.~V., {Pelt}, J., \& {Tuominen}, I. 2002, \aap, 394, 505

\bibitem[{{Berdyugina} \& {Usoskin}(2003)}]{berdyugina2003}
{Berdyugina}, S.~V., \& {Usoskin}, I.~G. 2003, \aap, 405, 1121

\bibitem[{{Brewer} \& {Fischer}(2018)}]{brewer2018}
{Brewer}, J.~M., \& {Fischer}, D.~A. 2018, \apjs, 237, 38

\bibitem[{{Cameron} \& {Sch{\"u}ssler}(2007)}]{cameron2007}
{Cameron}, R., \& {Sch{\"u}ssler}, M. 2007, \apj, 659, 801

\bibitem[{{Claret}(2000)}]{claret}
{Claret}, A. 2000, \aap, 363, 1081

\bibitem[{{Cox} \& {Pilachowski}(2000)}]{cox2000}
{Cox}, A.~N., \& {Pilachowski}, C.~A. 2000, Physics Today, 53, 77

\bibitem[{{Davenport} {et~al.}(2015){Davenport}, {Hebb}, \&
  {Hawley}}]{gj12432015}
{Davenport}, J. R.~A., {Hebb}, L., \& {Hawley}, S.~L. 2015, \apj, 806, 212

\bibitem[{{Donati} {et~al.}(1997){Donati}, {Semel}, {Carter}, {Rees}, \&
  {Collier Cameron}}]{donati1997}
{Donati}, J.~F., {Semel}, M., {Carter}, B.~D., {Rees}, D.~E., \& {Collier
  Cameron}, A. 1997, \mnras, 291, 658

\bibitem[{{Ducrot} {et~al.}(2018){Ducrot}, {Sestovic}, {Morris}, {Gillon},
  {Triaud}, {De Wit}, {Thimmarayappa}, {Agol}, {Almleaky}, {Burdanov},
  {Burgasser}, {Delrez}, {Demory}, {Jehin}, {Leconte}, {McCormac}, {Murray},
  {Queloz}, {Selsis}, {Thompson}, \& {Van Grootel}}]{Ducrot2018}
{Ducrot}, E., {Sestovic}, M., {Morris}, B.~M., {et~al.} 2018, The Astronomical
  Journal, 156, 218

\bibitem[{Foreman-Mackey(2016)}]{corner}
Foreman-Mackey, D. 2016, The Journal of Open Source Software, 1, 24.
\newblock \url{https://doi.org/10.21105/joss.00024}

\bibitem[{{Foreman-Mackey} {et~al.}(2013){Foreman-Mackey}, {Hogg}, {Lang}, \&
  {Goodman}}]{foreman2013}
{Foreman-Mackey}, D., {Hogg}, D.~W., {Lang}, D., \& {Goodman}, J. 2013, \pasp,
  125, 306

\bibitem[{{Gressier} {et~al.}(2022){Gressier}, {Mori}, {Changeat}, {Edwards},
  {Beaulieu}, {Marcq}, \& {Charnay}}]{Gressier2022}
{Gressier}, A., {Mori}, M., {Changeat}, Q., {et~al.} 2022, Astronomy and
  Astrophysics, 658, A133

\bibitem[{{Hathaway} \& {Choudhary}(2008)}]{hathaway2008}
{Hathaway}, D.~H., \& {Choudhary}, D.~P. 2008, \solphys, 250, 269

\bibitem[{{Howard} {et~al.}(1984){Howard}, {Gilman}, \& {Gilman}}]{howard1984}
{Howard}, R., {Gilman}, P.~I., \& {Gilman}, P.~A. 1984, \apj, 283, 373

\bibitem[{{Jetsu} {et~al.}(1993){Jetsu}, {Pelt}, \& {Tuominen}}]{jetsu1993}
{Jetsu}, L., {Pelt}, J., \& {Tuominen}, I. 1993, \aap, 278, 449

\bibitem[{{Kim} \& {Demarque}(1996)}]{kim1996}
{Kim}, Y.-C., \& {Demarque}, P. 1996, \apj, 457, 340

\bibitem[{{Kreidberg}(2015)}]{kreidberg2015}
{Kreidberg}, L. 2015, \pasp, 127, 1161

\bibitem[{{Kron}(1952)}]{kron1952}
{Kron}, G.~E. 1952, \apj, 115, 301

\bibitem[{{Lillo-Box} {et~al.}(2015){Lillo-Box}, {Barrado}, {Mancini},
  {Henning}, {Figueira}, {Ciceri}, \& {Santos}}]{lillobox2015}
{Lillo-Box}, J., {Barrado}, D., {Mancini}, L., {et~al.} 2015, \aap, 576, A88

\bibitem[{{Mandel} \& {Agol}(2002)}]{mandel2002}
{Mandel}, K., \& {Agol}, E. 2002, \apjl, 580, L171

\bibitem[{{McQuillan} {et~al.}(2013){McQuillan}, {Aigrain}, \&
  {Mazeh}}]{mcquillan2013}
{McQuillan}, A., {Aigrain}, S., \& {Mazeh}, T. 2013, \mnras, 432, 1203

\bibitem[{{McQuillan} {et~al.}(2014){McQuillan}, {Mazeh}, \&
  {Aigrain}}]{mcquillan2014}
{McQuillan}, A., {Mazeh}, T., \& {Aigrain}, S. 2014, \apjs, 211, 24

\bibitem[{{Mittag} {et~al.}(2018){Mittag}, {Schmitt}, \&
  {Schr{\"o}der}}]{mittag2018}
{Mittag}, M., {Schmitt}, J.~H.~M.~M., \& {Schr{\"o}der}, K.~P. 2018, \aap, 618,
  A48

\bibitem[{{Morris} {et~al.}(2018{\natexlab{a}}){Morris}, {Agol}, {Davenport},
  \& {Hawley}}]{Morris2018c}
{Morris}, B.~M., {Agol}, E., {Davenport}, J. R.~A., \& {Hawley}, S.~L.
  2018{\natexlab{a}}, The Astrophysical Journal, 857, 39

\bibitem[{{Morris} {et~al.}(2018{\natexlab{b}}){Morris}, {Curtis}, {Douglas},
  {Hawley}, {Ag{\"u}eros}, {Bobra}, \& {Agol}}]{morris2018}
{Morris}, B.~M., {Curtis}, J.~L., {Douglas}, S.~T., {et~al.}
  2018{\natexlab{b}}, \aj, 156, 203

\bibitem[{{Morris} {et~al.}(2019){Morris}, {Curtis}, {Sakari}, {Hawley}, \&
  {Agol}}]{morris2019}
{Morris}, B.~M., {Curtis}, J.~L., {Sakari}, C., {Hawley}, S.~L., \& {Agol}, E.
  2019, \aj, 158, 101

\bibitem[{{Morris} {et~al.}(2017){Morris}, {Hawley}, {Hebb}, {Sakari},
  {Davenport}, {Isaacson}, {Howard}, {Montet}, \& {Agol}}]{morris2017}
{Morris}, B.~M., {Hawley}, S.~L., {Hebb}, L., {et~al.} 2017, \apj, 848, 58

\bibitem[{{Morris} {et~al.}(2018{\natexlab{c}}){Morris}, {Agol}, {Hebb},
  {Hawley}, {Gillon}, {Ducrot}, {Delrez}, {Ingalls}, \& {Demory}}]{Morris2018b}
{Morris}, B.~M., {Agol}, E., {Hebb}, L., {et~al.} 2018{\natexlab{c}}, The
  Astrophysical Journal, 863, L32

\bibitem[{{Netto} \& {Valio}(2020)}]{netto2020}
{Netto}, Y., \& {Valio}, A. 2020, \aap, 635, A78

\bibitem[{{Newton}(1955)}]{newton1955}
{Newton}, H.~W. 1955, Vistas in Astronomy, 1, 666

\bibitem[{{O'Neal} {et~al.}(1998){O'Neal}, {Neff}, \& {Saar}}]{oneal1998}
{O'Neal}, D., {Neff}, J.~E., \& {Saar}, S.~H. 1998, \apj, 507, 919

\bibitem[{{O'Neal} {et~al.}(2004){O'Neal}, {Neff}, {Saar}, \&
  {Cuntz}}]{oneal2004}
{O'Neal}, D., {Neff}, J.~E., {Saar}, S.~H., \& {Cuntz}, M. 2004, \aj, 128, 1802

\bibitem[{{O'Neal} {et~al.}(2001){O'Neal}, {Neff}, {Saar}, \&
  {Mines}}]{oneal2001}
{O'Neal}, D., {Neff}, J.~E., {Saar}, S.~H., \& {Mines}, J.~K. 2001, \aj, 122,
  1954

\bibitem[{{O'Neal} {et~al.}(1996){O'Neal}, {Saar}, \& {Neff}}]{oneal1996}
{O'Neal}, D., {Saar}, S.~H., \& {Neff}, J.~E. 1996, \apj, 463, 766

\bibitem[{{Pollacco} {et~al.}(2008){Pollacco}, {Skillen}, {Collier Cameron},
  {Loeillet}, {Stempels}, {Bouchy}, {Gibson}, {Hebb}, {H{\'e}brard}, {Joshi},
  {McDonald}, {Smalley}, {Smith}, {Street}, {Udry}, {West}, {Wilson},
  {Wheatley}, {Aigrain}, {Alsubai}, {Benn}, {Bruce}, {Christian}, {Clarkson},
  {Enoch}, {Evans}, {Fitzsimmons}, {Haswell}, {Hellier}, {Hickey}, {Hodgkin},
  {Horne}, {Hrudkov{\'a}}, {Irwin}, {Kane}, {Keenan}, {Lister}, {Maxted},
  {Mayor}, {Moutou}, {Norton}, {Osborne}, {Parley}, {Pont}, {Queloz}, {Ryans},
  \& {Simpson}}]{pollacco2008}
{Pollacco}, D., {Skillen}, I., {Collier Cameron}, A., {et~al.} 2008, \mnras,
  385, 1576

\bibitem[{{Pont} {et~al.}(2007){Pont}, {Gilliland}, {Moutou}, {Charbonneau},
  {Bouchy}, {Brown}, {Mayor}, {Queloz}, {Santos}, \& {Udry}}]{pont2007}
{Pont}, F., {Gilliland}, R.~L., {Moutou}, C., {et~al.} 2007, \aap, 476, 1347

\bibitem[{{Quintana} {et~al.}(2021){Quintana}, {Col{\'o}n}, {Mosby},
  {Schlieder}, {Supsinskas}, {Karburn}, {Dotson}, {Greene}, {Hedges}, {Apai},
  {Barclay}, {Christiansen}, {Espinoza}, {Mullally}, {Gilbert}, {Hoffman},
  {Kostov}, {Lewis}, {Foote}, {Mason}, {Youngblood}, {Morris}, {Newton},
  {Pepper}, {Rackham}, {Rowe}, \& {Stevenson}}]{quintana2021}
{Quintana}, E.~V., {Col{\'o}n}, K.~D., {Mosby}, G., {et~al.} 2021, arXiv
  e-prints, arXiv:2108.06438

\bibitem[{{Rackham} {et~al.}(2018){Rackham}, {Apai}, \&
  {Giampapa}}]{rackham2018}
{Rackham}, B.~V., {Apai}, D., \& {Giampapa}, M.~S. 2018, \apj, 853, 122

\bibitem[{{Rackham} {et~al.}(2022){Rackham}, {Espinoza}, {Berdyugina},
  {Korhonen}, {MacDonald}, {Montet}, {Morris}, {Oshagh}, {Shapiro}, {Unruh},
  {Quintana}, {Zellem}, {Apai}, {Barclay}, {Barstow}, {Bruno}, {Carone},
  {Casewell}, {Cegla}, {Criscuoli}, {Fischer}, {Fournier}, {Giampapa}, {Giles},
  {Iyer}, {Kopp}, {Kostogryz}, {Krivova}, {Mallonn}, {McGruder},
  {Molaverdikhani}, {Newton}, {Panja}, {Peacock}, {Reardon}, {Roettenbacher},
  {Scandariato}, {Solanki}, {Stassun}, {Steiner}, {Stevenson}, {Tregloan-Reed},
  {Valio}, {Wedemeyer}, {Welbanks}, {Yu}, {Alam}, {Davenport}, {Deming},
  {Dong}, {Ducrot}, {Fisher}, {Gilbert}, {Kostov}, {L{\'o}pez-Morales}, {Line},
  {Mo{\v{c}}nik}, {Mullally}, {Paudel}, {Ribas}, \& {Valenti}}]{Rackham2022}
{Rackham}, B.~V., {Espinoza}, N., {Berdyugina}, S.~V., {et~al.} 2022, arXiv
  e-prints, arXiv:2201.09905

\bibitem[{{Santerne} {et~al.}(2012){Santerne}, {D{\'\i}az}, {Moutou}, {Bouchy},
  {H{\'e}brard}, {Almenara}, {Bonomo}, {Deleuil}, \& {Santos}}]{santerne2012}
{Santerne}, A., {D{\'\i}az}, R.~F., {Moutou}, C., {et~al.} 2012, \aap, 545, A76

\bibitem[{{Skumanich}(1972)}]{skumanich1972}
{Skumanich}, A. 1972, \apj, 171, 565

\bibitem[{{Solanki}(2003)}]{solanki2003}
{Solanki}, S.~K. 2003, \aapr, 11, 153

\bibitem[{{Suto} {et~al.}(2022){Suto}, {Sasaki}, {Nakagawa}, \&
  {Benomar}}]{suto2022}
{Suto}, Y., {Sasaki}, S., {Nakagawa}, Y., \& {Benomar}, O. 2022, arXiv
  e-prints, arXiv:2205.04836

\bibitem[{{Thompson} {et~al.}(2018){Thompson}, {Coughlin}, {Hoffman},
  {Mullally}, {Christiansen}, {Burke}, {Bryson}, {Batalha}, {Haas},
  {Catanzarite}, {Rowe}, {Barentsen}, {Caldwell}, {Clarke}, {Jenkins}, {Li},
  {Latham}, {Lissauer}, {Mathur}, {Morris}, {Seader}, {Smith}, {Klaus},
  {Twicken}, {Van Cleve}, {Wohler}, {Akeson}, {Ciardi}, {Cochran}, {Henze},
  {Howell}, {Huber}, {Pr{\v{s}}a}, {Ram{\'\i}rez}, {Morton}, {Barclay},
  {Campbell}, {Chaplin}, {Charbonneau}, {Christensen-Dalsgaard}, {Dotson},
  {Doyle}, {Dunham}, {Dupree}, {Ford}, {Geary}, {Girouard}, {Isaacson},
  {Kjeldsen}, {Quintana}, {Ragozzine}, {Shabram}, {Shporer}, {Silva Aguirre},
  {Steffen}, {Still}, {Tenenbaum}, {Welsh}, {Wolfgang}, {Zamudio}, {Koch}, \&
  {Borucki}}]{thompson2018}
{Thompson}, S.~E., {Coughlin}, J.~L., {Hoffman}, K., {et~al.} 2018, \apjs, 235,
  38

\bibitem[{{Torres} {et~al.}(2010){Torres}, {Andersen}, \&
  {Gim{\'e}nez}}]{torres2010}
{Torres}, G., {Andersen}, J., \& {Gim{\'e}nez}, A. 2010, \aapr, 18, 67

\bibitem[{{Vogt} {et~al.}(1987){Vogt}, {Penrod}, \& {Hatzes}}]{vogt1987}
{Vogt}, S.~S., {Penrod}, G.~D., \& {Hatzes}, A.~P. 1987, \apj, 321, 496

\bibitem[{{Wakeford} {et~al.}(2019){Wakeford}, {Lewis}, {Fowler}, {Bruno},
  {Wilson}, {Moran}, {Valenti}, {Batalha}, {Filippazzo}, {Bourrier},
  {H{\"o}rst}, {Lederer}, \& {de Wit}}]{Wakeford2019}
{Wakeford}, H.~R., {Lewis}, N.~K., {Fowler}, J., {et~al.} 2019, The
  Astronomical Journal, 157, 11

\bibitem[{{Wolter} {et~al.}(2009){Wolter}, {Schmitt}, {Huber}, {Czesla},
  {M{\"u}ller}, {Guenther}, \& {Hatzes}}]{wolter2009}
{Wolter}, U., {Schmitt}, J.~H.~M.~M., {Huber}, K.~F., {et~al.} 2009, \aap, 504,
  561

\bibitem[{{Zechmeister} \& {K{\"u}rster}(2009)}]{lombscargle}
{Zechmeister}, M., \& {K{\"u}rster}, M. 2009, \aap, 496, 577

\bibitem[{{Zirin}(1998)}]{zirin1998}
{Zirin}, H. 1998, {The Astrophysics of the Sun}

\end{thebibliography}

\end{document}